\documentclass[aps,rmp,reprint,amsmath,amssymb,shortbibliography,floatfix]{revtex4-1}
\usepackage{graphicx}
\usepackage{color}
\usepackage{hyperref}
\newcommand{\gar}{\alpha^\star}
\newcommand{\p}{\prime}
\newcommand{\K}{K$^+$ }
\newcommand{\Cl}{Cl$^-$ }
\newcommand{\Na}{Na$^+$ }
\newcommand{\Km}{\mathrm{K}}
\newcommand{\Clm}{\mathrm{Cl}}
\newcommand{\kt}{k_B T}

\newcommand{\qg}{\mathcal{G}}
\newcommand{\RH}{R_\mathrm{MH}}
\newcommand{\he}{h_p^\mathrm{eff}}
\newcommand{\pore}{\mathrm{pore}}
\newcommand{\access}{\mathrm{access}}

\newcommand{\bulk}{\mathrm{bulk}}
\newcommand{\surface}{\mathrm{surface}}
\newcommand{\volume}{\mathrm{volume}}
\newcommand{\LD}{\lambda_\mathrm{Du}}
\newcommand{\pH}{\textsl{p}H~}

\begin{document}

\title{Colloquium: Ionic phenomena in nanoscale pores through 2D materials}

\author{Subin Sahu}
\affiliation{\mbox{Biophysics Group, Microsystems and Nanotechnology Division, Physical Measurement Laboratory}, \mbox{National Institute of Standards and Technology, Gaithersburg, Maryland 20899, USA}}
\affiliation{\mbox{Maryland NanoCenter, University of Maryland, College Park, Maryland 20742, USA}}
\author{Michael Zwolak}
\email[\textbf{Corresponding author:} ]{mpz@nist.gov}
\affiliation{\mbox{Biophysics Group, Microsystems and Nanotechnology Division, Physical Measurement Laboratory}, \mbox{National Institute of Standards and Technology, Gaithersburg, Maryland 20899, USA}}

\begin{abstract}
Ion transport through nanopores permeates through many areas of science and technology, from cell behavior to sensing and separation to catalysis and batteries. Two-dimensional materials, such as graphene, molybdenum disulfide (MoS$_2$), and hexagonal boron nitride (hBN), are recent additions to these fields. Low-dimensional materials present new opportunities to develop filtration, sensing, and power technologies, encompassing ion exclusion membranes, DNA sequencing, single molecule detection, osmotic power generation, and beyond. Moreover, the physics of ionic transport through pores and constrictions within these materials is a distinct realm of competing many-particle interactions (e.g., solvation/dehydration, electrostatic blockade, hydrogen bond dynamics) and confinement. This opens up alternative routes to creating biomimetic pores and may even give analogues of quantum phenomena, such as quantized conductance, in the classical domain. These prospects make membranes of 2D materials -- i.e., 2D membranes -- fascinating. We will discuss the physics and applications of ionic transport through nanopores in 2D membranes.
\end{abstract}

\maketitle
\tableofcontents


\section{Introduction}
When the first strand of DNA was pulled through a biological ion channel~\cite{Kasianowicz1996-1}, a major, decades-long effort began to use ion transport -- and porous systems more generally -- for sequencing and molecular detection. After the isolation of graphene~\cite{novoselov2004electric} and the subsequent 2D tsunami, graphene and other materials joined this effort, becoming  2D membranes. Their unique electronic, chemical, and structural properties~\cite{geim2010rise} offer potential advantages over their biological and traditional solid-state counterparts in numerous applications. Graphene, for instance, is single atom thick and flexible but still mechanically robust. In pristine form, it is impermeable even to gases as small as helium~\cite{bunch2008impermeable} and is also an excellent ionic insulator~\cite{garaj2010}. Defects can be introduced to create pores of a controlled size that can selectively allow passage of certain gases, ions, or molecules. Ion transport through such a pore reveals physics at the atomic scale. The possibilities here become even more fascinating when considering that graphene should be amenable to a broad range of synthetic functionalization due to its carbon makeup.  

Moreover, 2D membranes have considerable potential in biosensing technologies. Their atomic thickness naturally gives spatial resolution at the molecular scale for detecting DNA nucleotides or other biomolecules. Both pores and channels provide  opportunities for measuring ion dehydration and its interplay with charge and functional groups. In addition, 2D membranes have become front and center as a candidate for filtration and selective transport. These include proposals for, and experiments on, novel desalination, gas separation, battery, and osmotic power technologies, among others.

Since this is a Colloquium, we do not give just a general review, listing topic after topic from the field. Rather, we aim to synthesize the myriad of results in the literature and deliver a firm foundation for ``new recruits'' and future progress, providing our perspective where appropriate. The very organization and content of this Colloquium are influenced by that perspective. We first cover the types of pores and channels (Sec.~\ref{Sec:Pores}), focusing heavily on biological ion channels and fabrication. Fabrication is the pillar of synthetic pore/channel research (and nanofluidics more generally). Biological channels are the paradigmatic ``advanced technology'', the ones we want to understand (via synthetic prototypes) and emulate (in applications). After setting this groundwork, we delve into the bulk of the review, the physics of ion transport, both continuum -- ``single body''-- (Sec.~\ref{sec:IonTransport}) and many-body (Sec.~\ref{sec:Interactions}). In these sections, we discuss the implications for applications (filtration and sensing) and fundamentals (biomimetic pores, measuring atomic-scale phenomena such as hydration and interactions), as well as simulation. We then briefly overview the technologies these membranes may enable (Sec.~\ref{sec:Apps}), tying back to the physics in prior sections. While we do not cover all potential applications, nor all experiments or proposals even when they fall within the purview of the Colloquium, we hope that readers will come away with the core knowledge of 2D membranes and their technological scope. We conclude with a synopsis of the field, future directions, and what we believe lies on the horizon (Sec.~\ref{sec:conc}).


\section{Nanopores and channels \label{Sec:Pores}}

Before exploring transport through pores in 2D membranes, it is essential to understand their predecessors -- biological ion channels and other solid-state pores -- and parallel developments, which set the context and scope of 2D membranes. We first give an overview of the different classes of pores, to which we dedicate quite some space as we hope it will provide an appreciation of where 2D membranes fit into the bigger picture and where they may help advance fundamental science and technology.

\subsection{Classes of nanopores \label{sec:Classes}}
Many types of nanoscale pores and porous systems are prevalent in nature. The most prominent among them are biological channels, which regulate the motion of ions and molecules across the cell membrane. These inspired the construction of artificial pores in solid-state membranes such as silicon nitride and silicon dioxide, which ultimately led to pores in 2D materials. There are, of course, numerous other examples of porous systems, such as zeolites and materials for batteries and separation. Some discussion will touch on aspects relevant to other examples, but our primary focus will be on pores in 2D membranes -- the advantages they convey and the groundbreaking applications they may enable. We thus start with a background on the behavior -- and fabrication -- of isolated pores, ones that led to the interest in 2D membranes. This background is intimately entwined with nanopore-based DNA sequencing. We, therefore, discuss the classes of pores mostly within this context.

\begin{figure}
\includegraphics{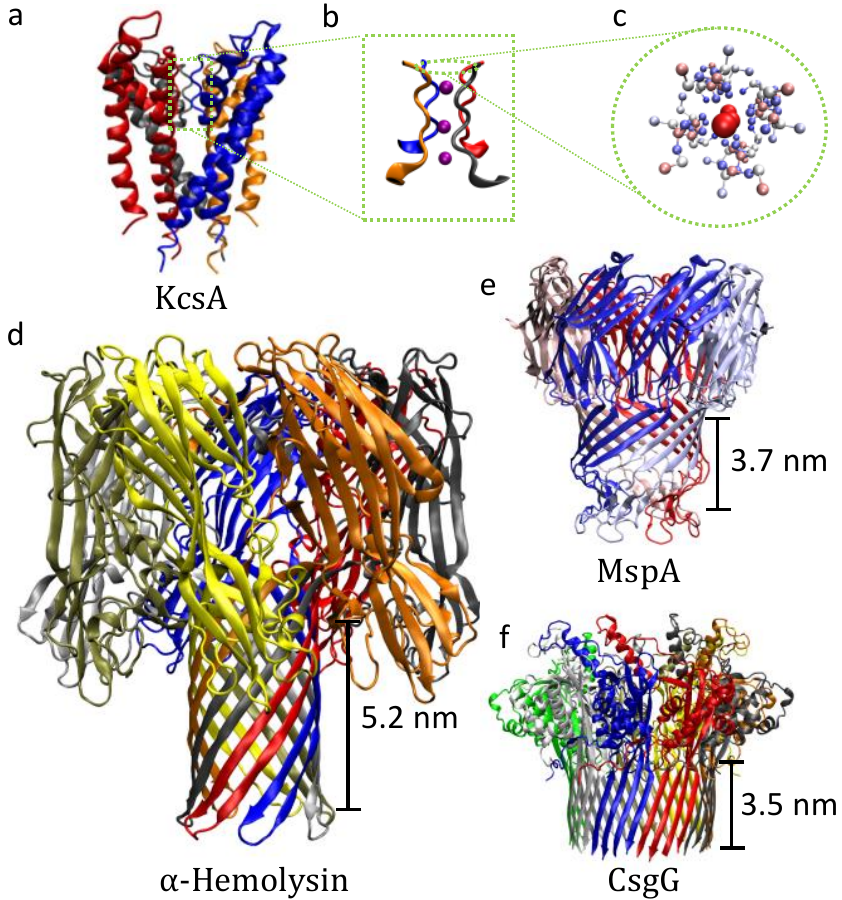}
\caption{Examples of biological ion channels. (a) The well-known potassium-selective channel KcsA. (b) Enlarged view of its selectivity filter with translocating \K ions (purple). (c) Top view of the selectivity filter. Colors indicate the atom charge from red (positive) to white (neutral) to blue (negative). (d-f) Various biological pores for  DNA sequencing studies. The length of the $\beta$-barrel -- the approximate sensing region -- is next to each channel. Shorter sensing regions are more successful in sequencing due to their higher spatial resolution. Colors indicate individual protein subunits.\label{fig:biopores}}
\end{figure}

\subsubsection{Biological ion channels \label{sec:BioChannels}}
Ion channels are membrane-spanning proteins that self-assemble into the lipid bilayer separating the cell from its environment~\cite{hille2001,zheng2015handbook}. These pores are present in all excitable cells, passively allowing ions to cross the cell membrane in the direction of the electrochemical gradient. This is in contrast to other membrane proteins, such as ion pumps and coupled transporters, which actively transport ions via work performed by ATP (adenosine triphosphate) hydrolysis, e.g., in driving a conformation change that pumps ions up a potential barrier~\cite{gadsby2009ion} or rely on opposing movement of another species, i.e., the coupled ``cross-transport'' of different ions~\cite{gadsby2009ion}. 

These channels play a vital role in many physiological functions including neurotransmission, hormone secretion, vision, muscle excitation, and the cardiac cycle. In the words of Clay Armstrong, ``Ion channels are involved in every thought, every perception, every movement, every heartbeat. They developed early in evolution, probably in the service of basic cellular tasks like energy production and osmotic stabilization of cells, and evolved to underlie the elaborate electrical system that provides rapid perception and control'' \cite{hille1999ion}.  

Ion channels are ``built'' on modular themes~\cite{ashcroft2006}; families of channels are each composed of identical or similar functional core, such as the selectivity filter (see Fig~\ref{fig:biopores}). Even so, mutation and malfunction of these channels can occur, resulting in diseases such as epilepsy, cystic fibrosis, arrhythmia, paralysis, among many others collectively called channelopathies~\cite{ackerman1997ion, cooper1999ion,catterall2010ion, ashcroft2006, ashcroft1999ion}. Delineating the different aspects of ion channel operation is thus one of the central motivations behind studying transport through pores, as it gives routes to designing corrective drugs and therapeutics~\citep{ackerman1997ion, hubner2002ion, catterall2010ion, bagal2012}.

Via their functional elements, ion channels act as the ``gatekeepers'' of the cell,  determining when and what gets through the cell membrane. These pores open and close -- i.e., {\em gate} -- in response to internal and external stimuli, 
such as ligand binding~\cite{brejc2001crystal} and the presence of certain chemical species~\cite{levitan1994modulation, hinman2006trp}, \pH level~\cite{grunder2015biophysical}, heat~\cite{caterina1997capsaicin}, pressure~\cite{martinac1987pressure}, mechanical stress~\cite{sadoshima1997cellular}, magnetic field~\cite{walleczek1992pulsed}, electric field~\cite{seoh1996voltage}, and various electromagnetic waves~\cite{pall2013electromagnetic} including visible light \cite{govorunova2015natural}. Together with ion pumps, gating forms the very basis of the nervous system of living organisms.

Ion channels can also let certain ion species pass while effectively blocking others -- i.e., they are {\em selective}. This allows channels to maintain the proper balance of ions in and outside of cells, called cellular homeostasis, which is critical for cell vitality and higher level function~\cite{cooper2000cell}. Selectivity in biological pores can sometimes simply be based on size, such as in gap junction proteins~\cite{veenstra1996size, heyman2008hindered} which allow movement of ions and small molecules lighter than $\approx 1000$ Da ~\cite{kumar1996gap}. 

Selectivity is more often specialized and leads to very high rejection of some ions compared to others, even ones that are quite similar. The potassium channel from {\em Streptomyces lividans} (KcsA, Fig.~\ref{fig:biopores}a) is a remarkable example, selecting \K over the similar size \Na at about a ratio of $10^4$ to 1 and simultaneously allowing \K ions to flow at near the diffusion limit~\cite{hille2001, Doyle98-1,  kopec2018direct}. The fundamental mechanism came to light in 1998 with the first crystallographic structure of KcsA~\cite{Doyle98-1}. This demonstrated that the so-called selectivity filter -- the region responsible for selection -- is lined with polarized functional groups in a very particular arrangement; see Figs.~\ref{fig:biopores}b,c. This not only repels ions of opposite charge but also compensates for the dehydration of specific ions -- their loss of tightly bound water molecules when entering the subnanoscale channel/pore  -- thus giving rise to the large \K over \Na selectivity despite their identical charge and similar size. These characteristics are turned on their head for sensing: The current flowing can indicate what species are in the pore. Ion channels have thus attracted enormous interest in ``next-generation'' DNA sequencing and molecular detection. Albeit indirectly, it is here where the story of graphene and other 2D membranes starts. 

\citet{Kasianowicz1996-1} were the first to demonstrate that DNA can be ``threaded'' through a nanopore. They examined single-stranded DNA (ssDNA) and RNA (ssRNA) translocation through $\alpha$-hemolysin (Fig.~\ref{fig:biopores}d), suggesting that ``ionic blockade" events -- how much current is suppressed by the presence of particular species within the pore -- can be employed in sequencing. This protein pore was the subject of considerable prior research, in particular, on how to keep the channel open and stable~\cite{menestrina1986ionic, bezrukov1993current, kasianowicz1995protonation}. Moreover, its smallest aperture is about 1.4 nm in diameter, just above the width of a single nucleotide and thus in the range that may allow blockade levels to be used to sequence. This pioneering work demonstrated that ssDNA could indeed pass through the pore and give rise to blockade events, and showed that the DNA length can be detected. It did not take long to show that $\alpha$-hemolysin can differentiate homogeneous sequences of ssRNA~\cite{Akeson1999-1} and ssDNA~\cite{meller2000rapid}.

These studies, though, put the challenge of sequencing into perspective. Due to the small changes in ionic current, the translocation rate needs to be slow enough for the electronics to identify the nucleotide(s) present. For $\alpha$-hemolysin, the translocation rate is 1 $\mu$s to 10 $\mu$s per base at a 120 mV applied voltage~\cite{meller2000rapid, meller2001voltage}. For the changes in the blockade current levels, less than 10 pA~\cite{deamer2002characterization}, there are only about 60 ions in a microsecond from which to differentiate the signal. When actually sequencing and the blockade is due to a few bases, the changes in current are even smaller. Thus, megahertz-level measurements are already hitting the Poisson limit. State-of-the-art measurements typically reach 100 kHz levels (e.g., 250 kHz). However, a suitable bandwidth is heavily dependent on the details of the application (pore-analyte interactions, longer sensing regions that average over many nucleotides prohibiting individual base detection, etc.)

Fortunately, biological pores confer a significant advantage -- they have a precise atomic construction, one that can be engineered with synthetic biology.  This enables them to be modified and integrated with other biological ``machines'' and molecular components. Eventually, the dwell time was increased to several milliseconds by using enzymes  -- such as the Klenow fragment \cite{benner2007} or exonuclease \cite{hornblower2007} -- that interact with DNA and slow down its translocation. Further progress was made in controllably feeding each nucleotide into the pore ``in turn'' via a DNA polymerase~\cite{cockroft2008, cherf2012automated}. 

These advances by themselves, of course, do not yield all the essential pieces of a full sequencing approach. In particular, accurate base identification (or, as is typically the case, few bases, e.g., quadromer, identification) requires a short length limiting aperture, on the order of the spacing of DNA bases in ssDNA (about 0.6 nm). This is in addition to a small aperture width. The length of the sensing aperture in $\alpha$-hemolysin is about 5 nm (see Fig.~\ref{fig:biopores}d) -- many times the distance between bases in DNA. Despite the long sensing region,~\citet{clarke2009} were able to identify the total composition of A (adenine), G (guanine), C (cytosine), and T (thymine) bases in a strand of DNA. They used an exonuclease enzyme in solution to cleave DNA into individual nucleotides which were sensed by an $\alpha$-hemolysin pore with a bound adapter molecule -- a molecule that fits into the pore and helps regulate the translocation rate and improve the blockade level. However, sequencing was not possible because the exonuclease was free floating and just broke apart the DNA in solution. Furthermore, even if an exonuclease was bound nearby the pore mouth (to feed nucleotides into the pore), theoretical arguments suggest that diffusion of the cleaved nucleotides would exponentially decrease the reading accuracy with the DNA length~\cite{reiner2012effects}. There are other challenges, of course, depending on the exact technique, such as the stochastic nature of motion at the atomic scale that hinders, e.g., DNA from passing in a linear, base-after-base fashion (nucleotides can move backward or linger, etc.). The two issues described above, though, were the significant roadblocks initially faced in the ultimate goal to devise a physically-based approach for DNA sequencing~\cite{Zwolak08,branton2008potential}.

While it is possible to improve the discrimination in $\alpha$-hemolysin by mutating the sensing region~\cite{stoddart2009,stoddart2010multiple}, an alternative is to start with a pore with a shorter sensing region such as in Mycobacterium smegmatis porin A (MspA, Fig.~\ref{fig:biopores}e)~\cite{trias1994permeability, niederweis1999cloning}. It has a $\approx 1.2$ nm wide smallest aperture and a “funnel” structure, which gives a length of about 0.6 nm to this region. MspA can distinguish DNA bases in proof-of-principle experiments with higher fidelity than $\alpha$-hemolysin~\cite{butler2008,derrington2010}. Still, it does not preclude adjacent nucleotides from contributing to the ionic blockade as there is a 3 nm long region where the constriction is narrow (i.e., the $\beta$-barrel, the approximate sensing region). When used in sequencing, about four bases affect the blockade current~\cite{laszlo2014decoding,bhattacharya2016water}. Another biological channel that  recently came into the spotlight is curli specific genes G (CsgG, Fig.~\ref{fig:biopores}f)~\cite{goyal2014}. CsgG is in the latest version of a commercial nanopore sequencer~\cite{brown2016nanopore}. Although there have been challenges in sequencing quality~\cite{mikheyev2014first}, these technologies are undergoing rapid development, improving performance and accuracy~\cite{bayley2015nanopore,jain2015improved}. Demonstrations include point-of-care diagnostics, such as detecting pathogens [e.g., Ebola \cite{quick2016real}], and even whole human genome sequencing~\cite{jain2018nanopore}.

Nanopore-based sequencing is possible, as exemplified by biological ion channel-based techniques. Their advantages enabled this achievement. Specifically, their atomically precise construction -- while undergoing fluctuations -- gives a pore with known and engineerable characteristics. The ability to select from the plethora of ``tried and true'' biological machines, mutate them, and integrate them gives a sm\"{o}rg\r{a}sbord of opportunity for sensing and molecular processing, such as modifying the interaction of the channel with different analytes. However, there are still limitations. These techniques are slow and require redundancy. Achieving high throughput requires thousands of pores in parallel~\cite{jain2016oxford}. Moreover, for general molecular detection, they are not stable under a wide range of conditions (pH, temperature, etc.)\ and require modification~\cite{heerema2016}. While highly modular, biological channels do not easily ``fit'' into our typical device paradigm. This is not a disadvantage per se, but it does hinder our ability to ``tune'' the device, for which typical solid-state setups have key tunable parameters, such as pore thickness/radius, probe position, etc. These aspects can be changed in biological systems, but often not continuously, or limited to within a specific range, and some parameters are ill-defined.

\begin{figure}
\includegraphics{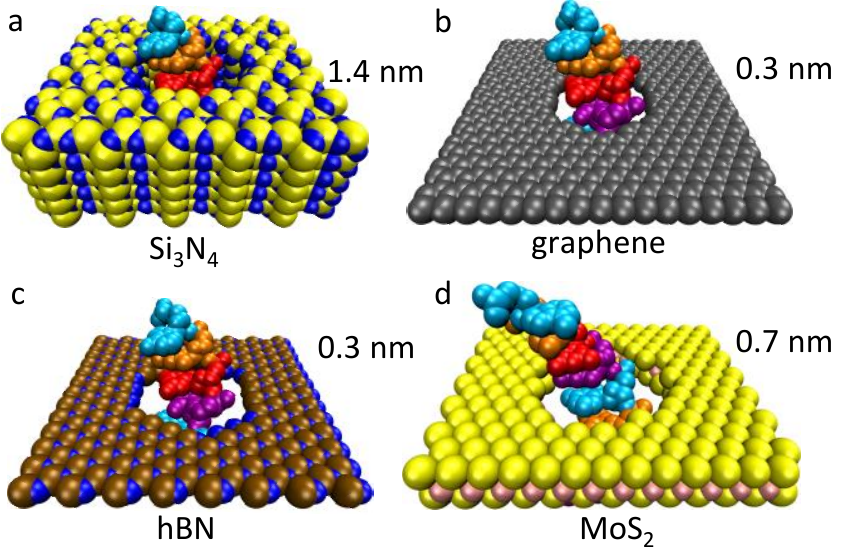}
\caption{Single-stranded DNA translocating through various pores. (a) SiN$_x$  (shown as Si$_3$N$_4$) pore at its minimum thickness (1.4 nm) so far achieved~\cite{rodriguez2015}. Almost all traditional solid-state membranes (including SiO$_2$ and other materials) are much thicker, giving pores 10 nm in length or longer. Three membranes with atomic or near-atomic thickness are (b) graphene, (c) hBN, and (d) MoS$_2$.\label{fig:SSpores}}
\end{figure}

\subsubsection{Solid-state nanopores \label{sec:SoldStatePores}}
Before the advent of ion channel approaches that met the core challenges above, the quest for rapid, low-cost sequencing generated tremendous interest in artificial pores in solid-state membranes, such as silicon nitride (SiN$_x$), silicon dioxide (SiO$_2$), polymers, and others \cite{ keyser2006direct, iqbal2007solid, dekker2007solid, branton2008potential,Zwolak08}. These pores can be more easily integrated with alternative probes, such as embedded electronics~\cite{Zwolak05,lagerqvist2006fast, lagerqvist2007influence, lagerqvist2007comment, krems2009effect, zwolak2012dna} or capacitive sensors~\cite{heng2005beyond, gracheva2006simulation}. While still under development, integration of nanoscale sensors may also revolutionize how we think about and perform molecular detection, including sequencing~\cite{Zwolak08}. These can potentially be operated at higher -- but still limited -- translocation rates due to larger currents (i.e., higher bandwidths). Other advantages of solid-state pores include the potential for manufacturing at a large scale (e.g., for ubiquitous sensing and sequencing), integration with solid-state electronic circuits for enhanced ionic current detection~\cite{rosenstein2012}, and operation in a broad range of conditions.

Fabrication of solid-state nanopores has seen significant progress over the last 20 years. Reactive ion etching~\cite{schmidt2000chip, fertig2000stable} and ion-track etching (ion bombardment followed by chemical etching)~\cite{fertig2001microstructured, siwy2002fabrication, siwy2003ion} give methods to create pores in thin silicon films. The channels formed by these chemical methods are rather large and asymmetric. To make smaller and more uniform pores, \citet{li2001ion} developed a technique that drills a hole in an ultra-thin membrane using a focused ion beam (FIB), called ion-beam sculpting. An ion sensor on the back side of the membrane provides feedback by measuring the total ion flux through the pore which scales with area, allowing for nanometer-scale control of the pore size. Additionally, the ion beam does not just eject matter but also facilitates the diffusion of surface atoms. Thus, by controlling intensity -- the rate of bombardment -- and temperature -- which determines the diffusion rate -- pores can be shrunk or expanded.

~\citet{storm2003fabrication} developed a method that uses a transmission electron microscope (TEM) to fine-tune the pores fabricated using other techniques such as chemical etching. They found that, when exposed to a wide-field TEM beam, large pores expanded whereas small pores shrank due to a surface tension effect. This allows pores to be controllably reduced in diameter while monitoring the TEM image. Alternatively, a focused TEM beam can also directly drill nanopores~\cite{heng2004sizing, krapf2006fabrication, kim2006}, which can be further refined with wide-field TEM~\cite{dekker2007solid}.

An orthogonal technique to create pores is dielectric breakdown~\cite{kwok2014nanopore}, which is inexpensive and more accessible since it does not require drilling with TEM or a FIB. In a standard nanopore setup, \citeauthor{kwok2014nanopore} applied a large electric field (1 V/nm) -- comparable to, but smaller than, the dielectric strength of the membrane material -- while monitoring the resulting tunnelling current through the membrane. This eventually opens a pore, determined from the sudden increase in current across the membrane. The pore is initially as small as 1 nm in diameter~\cite{briggs2015kinetics} and can be further enlarged with a moderate electric field, yielding subnanometer precision~\cite{beamish2012precise}. 
  
After the development of ion-beam sculpting and TEM approaches, several groups demonstrated that DNA molecules translocate through the solid-state nanopores (see  Fig.~\ref{fig:SSpores}a), and can be detected via the blockade current \cite{fologea2005detecting,li2001ion,meller2001voltage, li2003dna,storm2005translocation,storm2005fast}. Unfortunately, the two main problems that hindered early attempts of DNA sequencing via biological pores -- low temporal resolution due to fast translocation and low spatial resolution due to several bases being present in the sensing region simultaneously -- are worse in solid-state nanopores. Additionally, construction of these pores lacks the atomic precision provided by biological channels. The absence of control over the surface roughness and the charge distribution has severe implications for reproducibility (and gives additional noise). While differentiation of homopolymers has been achieved in solid-state pores~\cite{venta2013differentiation, akahori2017discrimination}, base-level discrimination has not been demonstrated, whether via the ionic current or embedded sensors~\cite{heerema2016}. Solid-state pores have been employed, though, to study fundamental aspects of polymer dynamics in confined geometries~\cite{chang2004dna, polonsky2007nanopore, luan2012slowing, belkin2013stretching, wang2014regulating}. Efforts continue to achieve sequencing, as such setups would be genuinely transformative, opening up a broad range of applications. This naturally leads us to 2D membranes.

\subsubsection{Atomically thin nanopores \label{sec:2Dpores}}

\begin{figure}
\includegraphics{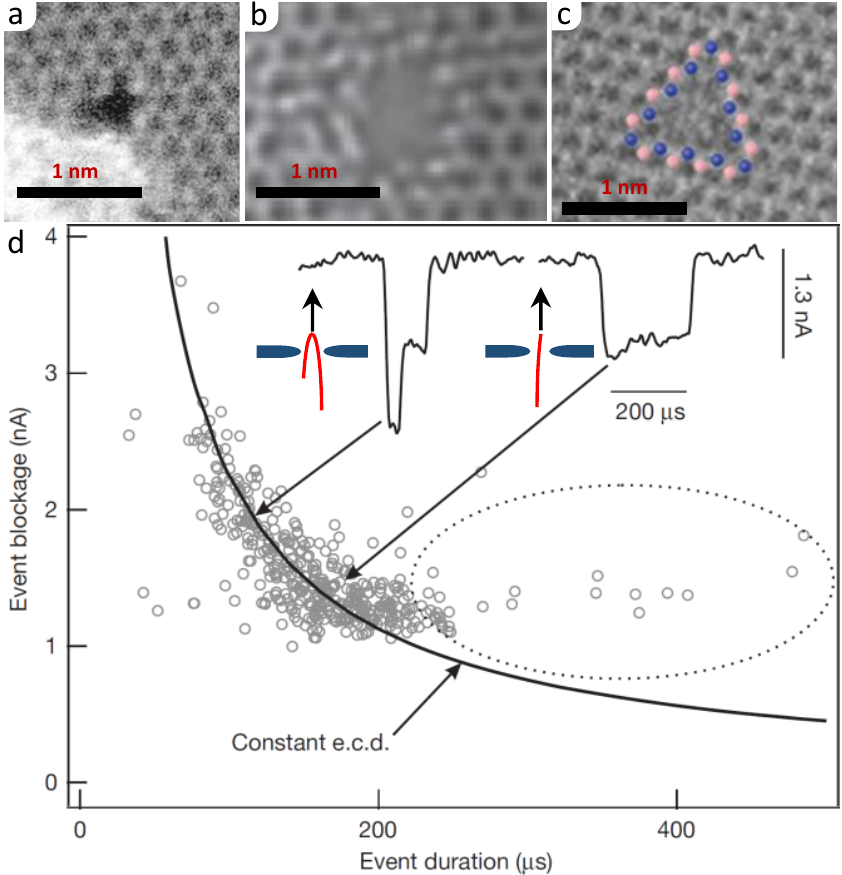}
\caption{Pores in 2D membranes. (a) Graphene pore of ``radius'' 0.19~nm fabricated via ion bombardment and chemical etching. From ~\citet{OHern2014}. (b) MoS$_2$ pore of ``radius'' 0.3~nm made via electrochemical breakdown. From ~\citet{feng2016observation}. (c) hBN pore from electron beam irradiation. From ~\citet{ryu2015atomic}. (d) Scatter plot of the  blockade current/duration for 10-kilobase dsDNA translocation through a graphene pore of diameter 5 nm.  The insets give events for partially folded (left) and unfolded (right) configurations. The electronic charge deficit (e.c.d.) indicates that, e.g., single folds block twice the charge but for half the time, giving a constant total blockade for the event. Adapted from \citet{garaj2010}. \label{fig:TEM2Dpores}}
\end{figure}

The isolation of graphene came at a time when researchers were exploring alternatives to biological ion channels for DNA sequencing. It was soon shown that these membranes could be sculpted with sub-nanometer scale precision~\cite{fischbein2008}. In fact, the fabrication of pores in 2D materials (Fig.~\ref{fig:SSpores}b-c) can be done in the same way as traditional solid-state membranes. To do so, a 2D material is suspended over a microscale hole in a substrate, such as SiN$_x$, and a nanoscale pore is drilled using a focused electron-beam in a TEM. The TEM, at lower energy, is also used to image the membrane and determine the size of the pore; see Figs.~\ref{fig:TEM2Dpores}a-c.

In 2010, DNA translocation through graphene nanopores was measured by three groups via the blockade current; see Fig.~\ref{fig:TEM2Dpores}d~\cite{merchant2010,garaj2010,Schneider2010}. These pores have the ``right'' thickness to potentially distinguish individual DNA bases, as it is similar to the distance between the consecutive bases (Fig.~\ref{fig:SSpores}). Hydrophobic effects, however, are a significant problem -- the nitrogenous bases of DNA molecules tend to stick to the nonpolar graphene as this reduces the contact surface with water. In addition to influencing the configurational dynamics of translocating DNA, such sticking can clog the pore, prohibiting further measurement or use. \citet{garaj2013} suggested that very high salt concentration allowed for the smooth translocation of double-stranded DNA (dsDNA) through the graphene pore; the effectiveness of this approach is debated nevertheless~\cite{schneider2013}. Coating graphene with a different material, such as pyrene ethylene glycol~\cite{schneider2013}, can prevent DNA from sticking, but this makes the membrane thicker and thus lower  spatial resolution. Another issue is the translocation rate -- when DNA does translocate through a graphene pore, it does so very fast. As mentioned earlier, this was also a significant issue in the biological case and was solved only after many attempts by several groups. Unfortunately, the solution for biological pores cannot be directly applied to these artificial pores, so researchers are trying different approaches to slow down the translocation rate, see Sec.~\ref{sec:Apps}.

Other 2D membranes, such as MoS$_2$~\cite{heiranian2015} and hBN~\cite{liu2013boron}, have also been studied for DNA sequencing. Encouragingly, \citet{feng2015identification} found that the problem of DNA sticking to the surface is reduced in MoS$_2$ due to hydrophilic Mo-rich clusters at the edge of the pore~\cite{liu2013top}. Similarly, hBN is also less hydrophobic compare to graphene and can be made more hydrophilic by UV-ozone treatment~\cite{zhou2013dna}. It is clear, as well, that 2D membranes offer other opportunities in sensing, such as using the in-plane electronic current to identify DNA bases~\cite{postma2010rapid,saha2011dna,girdhar2013graphene, traversi2013detecting,heerema2018probing} or using deflection to sense molecular binding or structural transitions~\cite{gruss2017communication,gruss2018graphene}. We  discuss these in Sec.~\ref{sec:Apps}.
\subsection{Pores in 2D membranes: Model ion channels?}

In addition to having the atomic resolution in the lateral direction, 2D membranes provide other advantages such as a highly ordered lattice that makes them mechanically robust~\cite{lee2008} and impermeable~\cite{bunch2008impermeable} despite their atomic thickness. While pores in 2D membranes can be formed more or less like traditional solid-state pores, they also give opportunities for nanoscale control and large-scale fabrication. 

\begin{figure}
\includegraphics{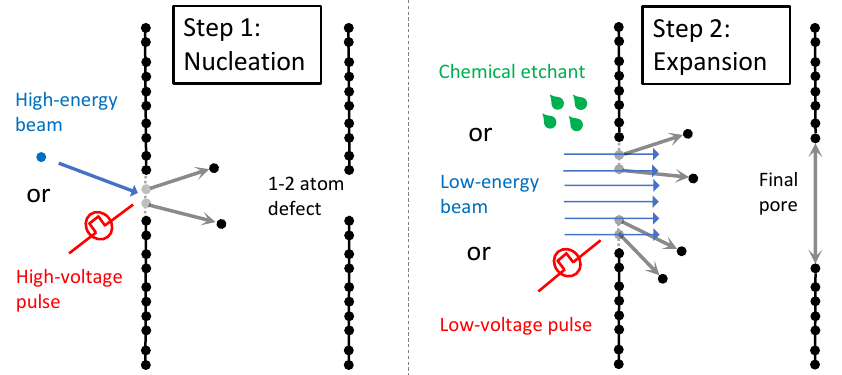}
\caption{``Atom-by-atom" techniques for graphene nanopore fabrication. In step 1, an ion beam~\cite{russo2012atom} or a high-voltage electric pulse~\cite{kwok2014nanopore,kuan2015electrical} creates a one- or two-atom defect in a suspended graphene sheet. In step 2, the defect expands to a pore by exposure to an electron beam, chemical etching (e.g., with KOH), or a low-voltage electric pulse. \label{fig:russogolovchenko}}
\end{figure}

For instance, an ``atom-by-atom'' technique employs energetic ions to create one to two atom defects in graphene, which are then slowly enlarged with an unfocused 80 keV electron beam; see Fig.~\ref{fig:russogolovchenko}~\cite{russo2012atom}. This selectively removes carbon atoms at the edge as their (estimated) $(14.1 \pm 0.1)$ eV displacement energy is below that of bulk carbons. The pore size is controlled via its linear growth rate. This technique works due to the atomic thickness of graphene; an ion beam cannot be used to drill an atom wide pore in traditional solid-state materials, but it can create defects like this in graphene at low enough intensities.

This general idea -- that as soon as a single carbon atom is ``knocked out'', a pore is nucleated -- also applies to other techniques, such ion bombardment followed by chemical etching~\cite{OHern2014}. The nucleation is the same, wherein defects are created using ion bombardment. However, chemical etching with KOH enlarges them to a size determined by the exposure time -- creating a relatively monodisperse set of pores -- eventually plateauing at a small value of the pore diameter. It was suggested that the termination of the pore edge by functional groups, such as ketone, quinone, hydroxyl, or carboxyl, could be stabilizing the pore. The related particle track etching, of course, can create pores in traditional membranes~\cite{apel2001diode, siwy2002fabrication, siwy2003ion}, such as SiN$_x$ and SiO$_2$, but the removal of one or two atoms is not enough to create the initial track; thus it lacks the atomic level control.

These thoughts apply across the board. Recently, \citet{kuan2015electrical} implemented the dielectric breakdown method for fabrication of pores in graphene.  For SiN$_x$ and SiO$_2$, there are slow changes and accumulation of defects which eventually results in a pore in $10^1$ s to $10^5$ s timescale depending on \pH and voltage. In graphene, however, pores ``nucleate'' with the removal of just a couple atoms. This happens rapidly, 250 ns voltage pulses already (stochastically) result in nucleation~\cite{kuan2015electrical}.  This is also seen in MoS$_2$ \cite{feng2015electrochemical}. While the enlargement process is done on the second timescale, the individual removal of atoms, or a couple of atoms, happens rapidly, well below the resolution of the measurement. The events, though, are separated enough in time that discrete steps in the ionic current are observed, giving a method of feedback control. Once again, this is due to the atomic thickness -- a pore does not need to span a thick layer of material. 

One might expect that this technique will result in many pores or a breakdown of the membrane. However, \citet{kuan2015electrical} show that this is not the case, as they obtain single pores, as small as 0.5 nm, with growth control of about 0.2 nm.   Given that the carbon bond length in graphene is 0.14 nm, this implies atomic control for pore enlargement. This also applies to MoS$_2$, as verified using TEM by \citet{feng2015electrochemical}. For MoS$_2$, pore formation likely starts at intrinsic defects that require lower energy for removal. Moreover, MoS$_2$ offers an additional advantage: The whole process can occur at quite low voltages (0.8 V, compared to 2.8 V for graphene \cite{feng2015electrochemical}, compared to 7 V pulse for graphene from~\citeauthor{kuan2015electrical}). \citeauthor{feng2015electrochemical} observe atomic steps (reflected in the ionic current) that occur over time, showing that they get essentially single atom removal -- the ``ultimate precision'' -- in the pore construction. Moreover, since the pores can be expanded using the same setup as the ion current measurement, this method allows for the study of multiple pore sizes using the same sample~\cite{rollings2016}, saving time and effort and removing some sources of sample-to-sample variation. 

What can these fabricated pores be used for? This is something that we will discuss throughout the Colloquium. However, we note that some techniques above enable the creation of pores or porous membranes with somewhat uniform pore sizes across a wide area. For filtration, desalination, etc., this is an ideal situation: Use the mechanically stable graphene membrane with a high concentration of pores of the same size to selectively let some species through (e.g., water) with minimum barrier, while blocking others (e.g., ions, organic molecules, etc.). Atomic precision allows one to tune the size, so it lets some species through ``fast'' but completely blocks others that are just a bit bigger. High flow rates require lots of pores but also a high permeability of individual pores, which graphene can provide.

There are, of course, still significant challenges. While the size is well controlled and there are potential approaches for large-scale fabrication, the precise characteristics of the pore (edge structure and pore/membrane functionalization) are not controlled or even known in some cases. Moreover, ~\citet{heerema20151f} showed that low-frequency ($1/f$ like) noise is dominant in graphene and hBN nanopores. Increasing the number of layers sharply decreased this noise, whereas ion concentration and \pH did not have a substantial effect. This, together with the presence of the noise for both graphene and hBN, suggests that it is due to mechanical fluctuations of the membrane that result in changes in both water structure and ion concentrations near the membrane and pore. As pointed out by \citet{kuan2015electrical} and \citet{heerema20151f}, the noise seems intrinsic to graphene and not the result of the fabrication process. However, further experiment and theoretical insight are necessary to confirm the origin and mechanism, whether mechanical or otherwise. As we discuss later, the application of strain to the graphene membrane may clearly delineate the role of mechanical fluctuations.

The ability to fabricate well-controlled-sized pores and uniform porous membranes are not the only advantages that 2D membranes offer. 2D membranes can be made with controllable and ``increasing'' (i.e., for systematic or specific studies) thickness by merely adding layers, in the spirit of 2D heterostructures discussed in other contexts~\cite{geim2013van}. As with other applications, it is imperative to both know and select for different layerings of graphene, e.g., monolayer over bilayer. This can be done both by optical means~\cite{meyer2007, blake2007} or by counting fringes at the edge of the layer~\cite{liu2009}. This control is genuinely at the atomic level, one to two to three, etc., atoms thick. Traditional solid-state membranes have a controllable thickness as well, including at nearly the atomic level~\cite{dekker2007solid}. However, this control is on top of an already thicker membrane; see Fig.~\ref{fig:SSpores}. The larger thickness affects flow rates, selectivity, and other relevant characteristics (not to mention uncontrollable surface characteristics, such as roughness/charges). 

Perhaps more intriguingly, these two types of controllability -- in effective diameter and length, both at the atomic level -- give possibilities for creating synthetic, biomimetic pores that exploit, quantify, and reveal the complex factors that contribute biological channel operation~\cite{sahu2017dehydration,sahu2017ionic}. The possibility to chemically functionalize graphene and other 2D membranes~\cite{hirunpinyopas2017desalination, lepoitevin2017functionalization} will open a vast phase space to create complex channels from the ground up. In addition to devising the proper chemistry for specific cases, the primary challenge is to selectively functionalize the pore edge only (or adhere multiple functional groups in a single pore), although even nonspecific functionalization has many potential uses in this regard (as well as technologically). This is the subject of Sec.~\ref{sec:Interactions}, where we discuss the basic physics of many-body transport. This follows a discussion of homogeneous, ohmic -- ``single-body'' -- transport in Sec.~\ref{sec:IonTransport}. We define ``many-body'' as the case where interactions, confinement, etc., become significant. This is not unlike the use of this term in quantum electron transport, except we have a purely classical system. 

Pores in 2D membranes are interesting because they can delineate properties of ion transport that are difficult or impossible to examine separately in biological or other solid-state systems. For example, the role of dehydration is hard to quantify in long pores due to its extreme sensitivity to the pore radius -- a small change in radius can exclude many water molecules, creating substantial energetic barriers and making currents undetectable. Since fractional dehydration is minimal in 2D pores (hydration layers can partially reside outside the pore while the ion is inside), a significant current can flow even as the pore size encroaches on the inner hydration. Thus the effects of dehydration, such as selectivity, can be directly probed/quantified. The dependence of access resistance on atomic factors can also be studied in 2D membranes. The prospects of 2D membranes in applications, such as molecular detection, biosensing, and filtration, make their study exciting but also requires a solid understanding of those contributions to ion transport. 


\section{Continuum ion transport}\label{sec:IonTransport}

At first glance, the description of ion transport through 2D membranes should be similar to the other channels and pores in Sec.~\ref{Sec:Pores}. However, while true, the atomic thickness and composition bring up a few notable differences: In contrast to nearly all other solid-state membranes, access resistance, rather than the pore resistance (both described below), is dominant for pores in 2D membranes with diameters above about 2 nm. When going to subnanometer pores, dehydration gives significantly smaller energy barriers in 2D membranes than in other solid-state systems. In this section, we will describe a typical approach to ion transport and highlight the differences for 2D membranes. 

\begin{figure}
\includegraphics{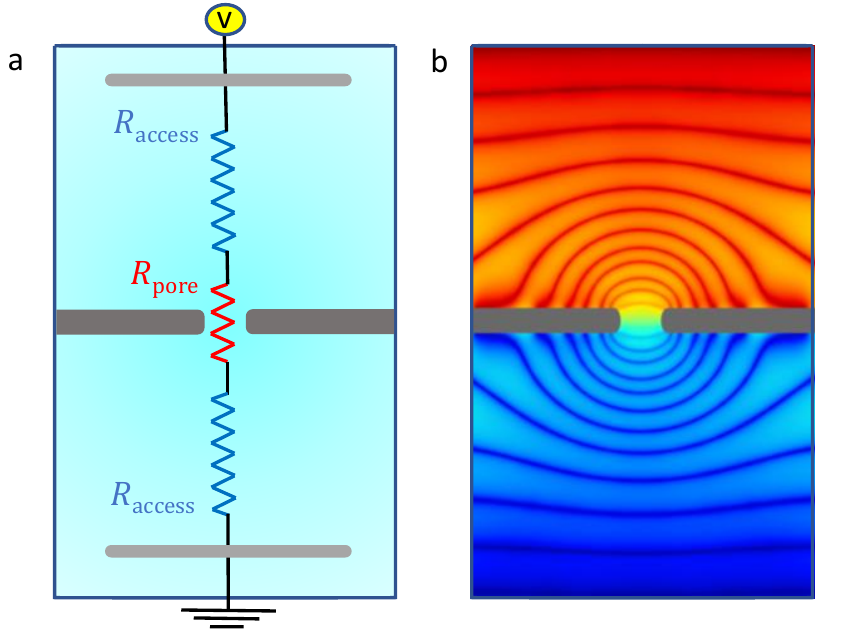}
\caption{Series representation of the ionic resistance. (a) The fluidic cell. The membrane (dark gray), whether composed of a 2D material, a traditional solid-state material, a biological membrane (e.g., lipid bilayer), or some combination (often a windowed SiN$_x$ membrane with a 2D material over top), separates two ionic solutions. An applied voltage (via two electrodes, light gray) across the membrane drives an ionic current through the pore. The equivalent circuit shows the access resistance ($R_{\mathrm{access}}$, blue resistors, equal for symmetric electrolytes) and the pore resistance ($R_{\mathrm{pore}}$, red resistor). (b) Equipotential surfaces from a continuum simulation. The access region develops hemispherical -- more accurately, spheroidal -- surfaces, essentially showing that the bulk converges  ``radially'' inward toward the pore. Within long, homogeneous pores with a symmetric electrolyte, flat potential surfaces develop and ions are flowing along the pore axis. This region is of ``negligible'' length in 2D membranes, creating an interesting competition between asymmetric electrolytes, imperfect geometries and fluctuations, dehydration, screening, and, potentially, functional groups. \label{fig:schematic}}
\end{figure}

Ion transport through a nanopore is equivalent to the current flowing through a circuit composed of a series of resistors as shown in Fig.~\ref{fig:schematic}a. A voltage bias (or an electrochemical potential gradient) from one side of the membrane to the other drives ions through the pore. The resistance for ions to transfer from one end of the pore to the other is the pore resistance. Conversely, the resistance for ions to converge from the bulk electrolyte away from the membrane to the mouth of the pore is the access resistance (variously known as the convergence resistance, interfacial resistance, contact resistance, and a component of the series resistance) and occurs on both sides of the membrane. Even though both resistances influence ion transport, pores in 2D membranes differ from those in other membranes in the balance of these two contributions. It is worth isolating this difference.

\subsection{Pore resistance \label{sec:PoreRes}}
The textbook pore resistance is associated with the current flowing uniformly through a region of cross-sectional area $A_p$ and constant electric field $E_p$,
\begin{equation} \label{eq:TypicalElectroDiff}
I = \sum_\nu q_\nu n^p_\nu \, \mu^p_\nu  E_p A_p,
\end{equation}
where $q_\nu$  is the charge, $n^p_{\nu}$ the concentration, and $\mu^p_\nu$ the mobility of ion species $\nu$ in the pore. The additional label $p$ indicates that these quantities can change inside a pore, especially when the pore is of nanoscale dimensions. For instance, there may be free-energy barriers (e.g., due to dehydration) or potential wells (e.g., due to favorable electrostatics) that change $n^p_\nu$ from its bulk value. The interaction of ions -- or their hydration layers -- with the pore walls (including functional groups) modifies $\mu^p_\nu$.

Assuming a constant potential drop, $E_p=V_p/h_p$, across the pore of length $h_p$, the pore resistance is 
\begin{equation} \label{eq:poreR}
R_p=\gamma_p \frac{h_p}{A_p},
\end{equation}
where we introduce the resistivity of the medium within the pore as $\gamma_p = 1/\sum_\nu q_\nu\,n^p_\nu\,\mu^p_\nu$. In practice, the pore resistance is much more complicated than the above equations indicate. The assumption of a uniform potential drop and cross-section, as well as a simplified contribution from ion-membrane interactions and functional groups/charges, do not hold in general. For instance, the different size of the ions (including hydration) creates nonuniformities in the potential, as some ion types can move closer to the membrane, and this ``bends'' the equipotential surfaces~\cite{sahu2018maxwell,sahu2018golden}. 

Biological pores and long solid-state pores, including even atomically thin pores, do not have uniform cross-sections. For fabricated pores, the drilling/etching processes introduce geometric and electrostatic (surface charge) nonuniformities. At the nanoscale, these cannot be described by average quantities, nor can one define the pore radius or accessible area for ion flow independent of the ionic species. These characteristics are {\em contextual} -- a term that will come up repeatedly in this Colloquium. In other words, they depend on multiple aspects of the setup. Other characteristics (pore length, charge, etc.) require similar considerations. Fluctuations, structural transitions, temperature, \pH, and so forth, can all influence primary pore and membrane characteristics. 

These issues will be addressed later. For now, however, we assume the simple picture expressed in Eqs.~\eqref{eq:TypicalElectroDiff}--\eqref{eq:poreR}. These do not capture everything, but they go a long way toward understanding ion transport and the general differences between 2D membranes and other pores. Going beyond this simple picture requires all-atom molecular dynamics (MD) simulation (or, at least, Brownian or Poisson-Nernst-Plank simulations), and thus introduces a higher level of complexity. We will, however, discuss how such simulations can be properly employed to address these additional complications. 

\subsection{Access resistance \label{sec:AccessRes}}
Access resistance is defined as the resistance for ions to converge from the bulk electrolyte to the mouth of the pore. This results in the spheroidal equipotential surfaces, directing ion flow inward toward the pore (Fig.~\ref{fig:schematic}b). Access resistance is fundamentally different than the pore resistance: It is the resistance of bulk medium rather than the pore itself (albeit, it is the resistance of the bulk medium ``in contact'' with the pore, and thus it is a property of both in concert). 

For a circular pore, the access resistance is 
\begin{equation} \label{eq:hall}
\RH=\frac{\gamma_b}{4a},
\end{equation}
where $\gamma_b$ is the resistivity of the bulk medium and $a$ is the pore radius. We denote this resistance $\RH$ where the `MH' is for Maxwell-Hall. While \citet{Hall1975} is normally credited with this equation for ion channels and pores, Maxwell already derived this form in the 1800's for electrical diffusion to an orifice~\cite{maxwell_1892}.  As noted above, access resistance goes by various names due to the variety of context in which it appears, e.g., thermal transport~\cite{GrayMathews1895,grober1921}, gas diffusion~\cite{brown1900}, and electrical contacts~\cite{Holm1958contact}. Any time there is a constriction, the normal bulk flow -- of anything, heat, gaseous particles, electrons, ions -- is interrupted, introducing a resistance.

Equation~\eqref{eq:hall} assumes that the medium is homogeneous with no concentration gradients or charge accumulation. It further assumes (i) a uniform potential at the mouth of the pore, (ii) no perpendicular electric field on the membrane, and (iii) a hemispherical electrode at infinity with a constant potential. However, the boundary condition (i) is almost never satisfied in ion transport, especially in biological ion channels where pore charges and functional groups give a strong coupling between the potential in the pore and its surroundings~\cite{luchinsky2009self}. Similarly, the presence of a membrane charge will alter the boundary condition (ii). The boundary condition (iii) is an idealization to simplify calculations: The electrodes are far away and the influence of the pore propagates outward radially, like the response of the homogeneous and isotropic medium to a point perturbation; and thus one can replace a distant disc electrode with a hemispherical one. This approximation, however, does not hold when the electrode(s) are close, such as in scanning ion-conductance microscopy~\cite{hansma1989}. Another factor that influences access resistance is  concentration polarization~\cite{kim2007concentration} due to selectivity: The preferred ion builds up on its exiting side, creating a field opposite to the applied field and making the access resistance voltage dependent~\cite{lauger1976,peskoff1988electrodiffusion}. Similarly, differences in mobility, size, charge, and electrostatic screening length between cation and anion will cause asymmetry in the equipotential surfaces and resistance on the two sides of the pore~\cite{sahu2018maxwell}. Despite these complications, the access resistance is expected to depend inversely on the pore radius. As with the pore resistance, additional (even contextual) complexities come in due to the presence of fixed charges, functional groups, and geometric variations.

\subsection{Total resistance \label{sec:TotalRes}}
Combining the pore and the access (on both sides of the membrane) contributions, the total resistance for cylindrical pore of radius $a$ and thickness $h_p$ is 
\begin{equation}\label{eq:TotalRes}
R=\gamma \left( \frac{1}{2a} + \frac{h_p}{\pi \, a^2} \right)
\end{equation}
where we take $\gamma=\gamma_p=\gamma_b$ for simplicity (along with the assumptions given in Secs.~\ref{sec:PoreRes} and~\ref{sec:AccessRes}, which we stress ignores dehydration and  interactions with charges and functional groups). This equation is often used for estimating pore size in experiments \cite{feng2015electrochemical} where direct measurement is difficult. Equation~\eqref{eq:TotalRes} is for the steady state. When a biomolecule translocates through the pore, dynamical effects can be present, such as the adjustment of the charge layers to the  resistance change~\cite{balijepalli2014}. These and stray capacitive effects contribute to high frequency noise. 

Equation~\eqref{eq:TotalRes} entails the fact that the relative contributions of pore and access resistance depend on the ratio of the pore thickness to its radius, $h_p/a$. Thus, it suggests that pore resistance will dominate the ion transport characteristics of biological ion channels and long solid-state pores when $h_p\gg a$. One can also create microscale pores~\cite{tsutsui2012} -- or just pores with diameters much larger than the membrane thickness~\cite{Kowalczyk2011} -- where access resistance dominates. For genuinely nanoscale pores, though, it is challenging to create membranes thin enough to tip the balance in favor of access resistance, although down to 1.4 nm thin membranes made of SiN$_x$ have been fabricated~\cite{rodriguez2015}. Nevertheless, the presence of surface charges and functional groups can drastically decrease the pore resistance, shifting this balance.
\begin{figure}
\includegraphics{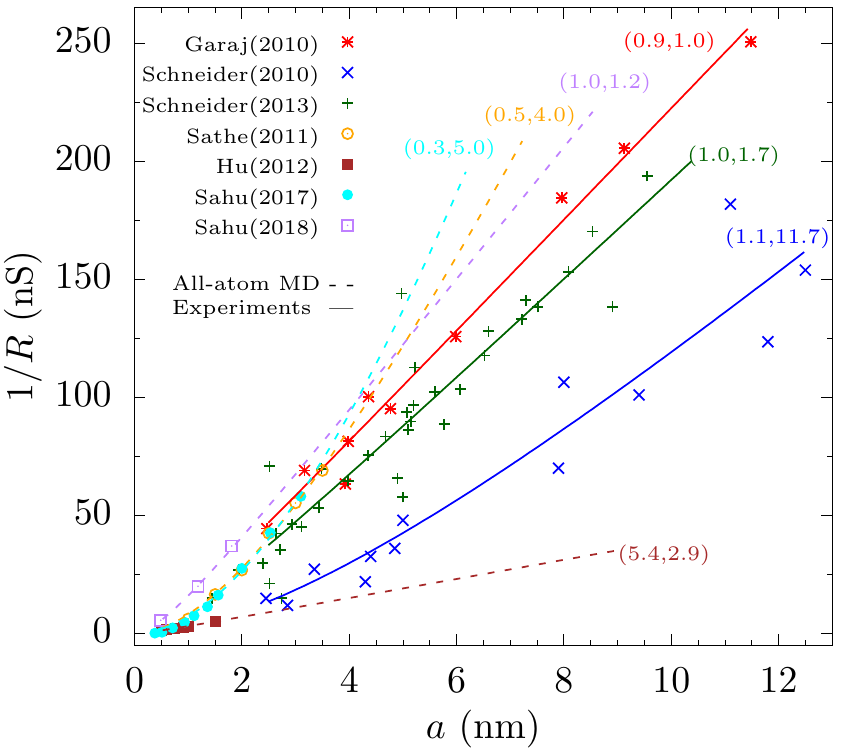}
\caption{Open pore conductance ($1/R$) versus pore radius $a$ in graphene from experiment and MD simulations. We fit the published data using Eq.~\ref{eq:Rfit} with $\lambda$ (the weight of access contribution) and $\he$ (the effective membrane thickness) as fitting parameters [shown as the pair $(\lambda,\he)$ next to the fitted lines. We use a resistivity of $\gamma=0.095\, \Omega $m  for experiments~\cite{garaj2010,Schneider2010,schneider2013} and MD with SPC/E water~\cite{Hu2012}, $\gamma=0.071\,\Omega $m for MD with TIP3P rigid water~\cite{sahu2018maxwell}, and $\gamma=0.081\,\Omega $m for MD with TIP3P flexible water~\cite{sathe2011,sahu2017dehydration}. Results from \citet{garaj2010} fit with the classical model with $\he \approx 1$ nm as expected for graphene membrane. The results from~\citet{Schneider2010} give an $\he$ 10 times larger, in part due to including many layer graphene pores (we note that the best fit gives both access and pore contributions, unlike their finding that it has only a pore contribution). However, their follow-up results~\cite{schneider2013} fit with the classical model and give an $\he$ consistent with other work. MD results by \citet{sathe2011} and \citet{sahu2017dehydration} give a small $\lambda$ and large $\he$. \citet{Hu2012} found a small conductance, see the text.  Recently, \citet{sahu2018maxwell} demonstrated that a finite-size scaling of the simulation cell and a pore-size correction accounting for hydration yield MD results in the classical form. The deviation of this result from experiment is solely due to the bulk conductivity given by MD. The fit errors are in the SM.\label{fig:GrapheneRes}}
\end{figure}

For 2D membranes, one might expect the pore resistance to be vanishingly small due to the small pore/membrane thickness $h_p$. However, for most applications, the desired pore size is also on the nanoscale ($<$ 10 nm); thus, the pore resistance can still be significant. The finite size of ions, including hydration, is also an essential factor in determining the accessible area (and thus $a$). As we will see, the effective thickness of monolayer graphene is $h_p\approx 1$  nm~\cite{garaj2010,sahu2018maxwell} instead of about 0.3 nm implied by the van der Waals (vdW) diameter of carbon atoms. This is because ions themselves have vdW diameters of about 0.3 nm and steric hindrance, also known as Born repulsion~\cite{sparreboom2009principles}, of the pore edge and the hydration layers also reduce the accessible area for transport; thus, the effective size of the pore is generally smaller than the size determined by the position of the edge atoms~\cite{sahu2018maxwell}. However, there is another competing factor that increases the accessible pore area: the flexibility of the membrane.  In contrast to long solid-state pores, graphene pores are more flexible, and their {\em dynamic area} can be larger than the static~\cite{sahu2018maxwell}. Additionally, structural fluctuation of the pore can both enhance (via induced flow) or decrease (via entropic trapping) diffusion, depending on the fluctuation frequency and the pore characteristics, such as the channel height ~\cite{marbach2018transport}. In 2D membranes, though, it seems less likely that fluctuations will hinder the translocation. Fluctuations in pore size, in particular, will tend to enhance transport via a skewed weighting of currents in the more open state.

The balance of pore and access resistance has been studied for graphene. Due to graphene's atomic thickness, the total resistance should vary as $1/a$ rather than $1/a^2$ for all but the smallest pores. However, early experiments were inconsistent, as were simulations. To make this more quantitative, we fit (assuming uncharged, unfunctionalized graphene) the resistance to
\begin{equation}
R=\gamma \left( \frac{\lambda}{2a} + \frac{\he}{\pi \, a^2} \right), \label{eq:Rfit}
\end{equation}
where $\lambda$ should be 1 according to Eq.~\eqref{eq:TotalRes} and $\he$ is the effective pore/membrane thickness. Equation~\eqref{eq:Rfit} is plotted in Fig.~\ref{fig:GrapheneRes} with $\lambda$ and $\he$ as fitting parameters, along with the experimental results and a suitable $\gamma$.  

Among the three original experimental papers on ion transport through graphene nanopores, one found a dominant $1/a$ behavior~\cite{garaj2010}, another $1/a^2$~\cite{Schneider2010}, and the third did not test the radius dependence (and also had higher currents and wide variation from device to device, which the authors contributed to the pinholes in the membrane)~\cite{merchant2010}. These disparate  results raised questions about the effective dimensionality of graphene at that time~\cite{siwy2010nanopores}. The second group later refined their fabrication technique (using a higher temperature of 600$^\circ$C), subsequently finding a dominant $1/a$ dependence~\cite{schneider2013} attributing the discrepancy with earlier results to amorphization of the pore edge by the electron beam at room temperature and (re)deposition of carbon and contaminants. This destroys the local crystal structure of graphene, creating uncertainty in pore height and geometry~\cite{xu2012size} and prohibits a proper assessment of the balance of access and pore resistance.

The computational results were also contradictory, which is troublesome since computational setups are generally well controlled. One study found $1/a^2$ behavior~\cite{sathe2011} and the right magnitude of the resistance (within the range of pore radii investigated), whereas another study found $1/a$~\cite{Hu2012}. The latter examined radii ($\le 1.5$ nm) in a regime where both pore and access contributions should be important, and the magnitude of the conductance was an order of magnitude less than experiment. The discrepancy may be due to the large fields employed or statistical uncertainties. Others have also reported a dominant $1/a^2$ dependence in simulation ($R_\access= 0.7\,\gamma/2a$ and $R_\pore= 6.7\,\gamma/\pi a^2$)~\cite{liang2013theoretical}. \citet{suk2014ion} cast their MD results in the form of Eq.~\eqref{eq:TotalRes} (i.e., $\lambda=1$) using a radius dependent conductivity. This approach uses a pore-size dependent conductivity in the access region. The access resistance, though, requires the bulk conductivity, as it is ions in the bulk which are converging towards the pore. The Supplementary Material (SM) has details of the fitting.

The overweighting of $1/a^2$, or obtaining results out of reasonable bounds, is thus perplexing. Computation does suffer from one major issue (besides general uncertainty in force fields): The limitations on simulations due to computational cost -- regarding both spatial cell size and timescales -- hinders the ability to capture the access resistance since it requires incorporating how the bulk converges to the pore. This convergence is algebraically decaying away from the pore and is thus quasi-long range~\cite{sahu2018maxwell}. Moreover, the resistance to ``normal'' bulk flow can be substantial in typical computational setups for, e.g.,  MD~\cite{sahu2018maxwell,sahu2018golden}, whereas it is negligible in experiments (less than 5 k$\Omega$~\cite{ho2005electrolytic} compared to typical resistances~\cite{hille1968, hamill1981improved} in the megaohm to gigaohm range) and not even considered. We stress that the ``normal'' bulk contribution (which depends on bulk dimensions and is independent of the pore/membrane) is distinct from the access contribution (which is a property of contact between the bulk and the pore and is independent of bulk dimensions in the infinite bulk limit). When the pore resistance is large (e.g., certain biological and solid-state nanopores), this access contribution can be negligible, but, for 2D membranes, it cannot be ignored, or incorrectly incorporated into the simulation, above the dehydration limit.

As with critical systems, e.g., extracting energy gaps and the decay of correlations \cite{fisher1972scaling,fisher1969decay}, a scaling analysis can adequately account for the normal bulk resistance and allow for the proper incorporation of the access contribution {\em for finite and small simulation cells}~\cite{sahu2018maxwell,sahu2018golden}. This analysis was developed in the context of graphene to resolve the computational discrepancy above and to shed light on issues that can arise in, or comparing to, experiment~\cite{sahu2018maxwell,sahu2018golden}. 

The equipotential surfaces, which are dictated by the spatial dependence of the resistance, are in Fig.~\ref{fig:schematic}b for a finite-size simulation cell of height $H$ and cross-sectional length $L$. These surfaces show the same behavior as an infinite cell up to a distance ($l\sim L/2 $) from the pore -- namely, spheroidal surfaces with ions converging inward toward the pore.  Taking the rest of the simulation cell to be composed of a normal bulk ``far'' from the pore, and a transition region between the two, accounts for the different dependencies of the resistance on artifacts of the simulation. For a finite-size cell of arbitrary dimensions (but $H>L$), this gives 
\begin{align}
R &= \frac{\gamma}{\qg} \left( \frac{H}{L^2}- \frac{\gar}{L} \right)+R_\infty = \frac{\gamma}{\qg L} \left( \alpha-\gar \right)+R_\infty, \label{eq:golden}
\end{align}
 where $\qg$ is a geometric constant ($\qg=1$ for rectangular cells and $\qg=\pi/4$ for cylindrical), $\gar=1.2$ is a constant for rectangular cells ($\gar=1.1$ for  cylindrical), and $\alpha=H/L$ is the aspect ratio~\cite{sahu2018golden}. The resistance, $R_\infty=(2 \RH +R_\pore)$, is for an infinite and balanced ($L\approx H \rightarrow \infty$) system and has the form of Eq.~\eqref{eq:TotalRes} under appropriate conditions but it will take on different forms for other conditions. This expression can rid simulation of normal bulk effects and capture the access resistance~\cite{sahu2018maxwell, sahu2018golden}. Nonetheless, it also suggests that it is best to use a simulation cell with a constant aspect ratio and do a finite-size scaling analysis. When $\alpha=\gar$, $R=R_\infty$ for any $L$. The constant $\gar$ is thus a special ratio -- the {\em golden aspect ratio} -- that removes finite-size effects. 

We can employ Eqs.~\eqref{eq:Rfit}-\eqref{eq:golden} to understand (and sometimes reanalyze when all data is available) prior simulations. For instance, \citet{suk2014ion} had aspect ratios in the range 0.9--1.2, close to the golden aspect ratio. Directly fitting their data for pore radii greater than 1 nm (above the dehydration threshold) gives $R_\access= 0.8\gamma/2a$ and $R_\pore= 2.0\,\gamma/\pi a^2$ (see the SM). This is only a mild overestimate of the pore over access resistance due to the proximity to $\gar$.

The caveats in the above scaling approach are that (i) the bulk resistivity, $\gamma$, from MD can differ from experiments, (ii) the effective membrane thickness is somewhat larger than expected, and (iii) the pore diameter is {\em contextual}. The caveat (i) is due to the inability of force fields to replicate the nonlinear behavior of $\gamma$ at high concentrations seen in experiments. Caveat (ii) is also not surprising; experiment yields an effective thickness of 0.6 nm (with error range 0 nm to 1.5 nm), found by fitting to finite element simulations~\cite{garaj2010,garaj2013}, which agrees with the theory to within error bars (note that in Fig.~\ref{fig:GrapheneRes} we fit Eq.~\eqref{eq:Rfit}, which gives about 1 nm effective thickness). The pore diameter, caveat (iii), is interesting. Clearly, even a symmetric pore is not perfectly circular -- it is not clear how to define the pore radius. In general, MD studies take the radius as the distance of the edge atoms from the pore center.  For large enough pores, minor geometric imperfections and the finite size of atomic species should not be significant, and the pore radius should be roughly just the radius of the opening. However, this is an issue for small pores, whether in experiment \cite{garaj2013, OHern2014, jain2015heterogeneous}  or theory~\cite{sahu2017dehydration, sahu2017ionic, sahu2018maxwell}. \citet{OHern2014}, for instance, defined the effective radius from the TEM imaged opening, which accounts at least for the finite-size of carbon atoms at the pore edge through their electron cloud. \citet{suk2014ion} defined the pore radius from the water density profile, which also captures electronic repulsion.

\begin{figure}
\includegraphics{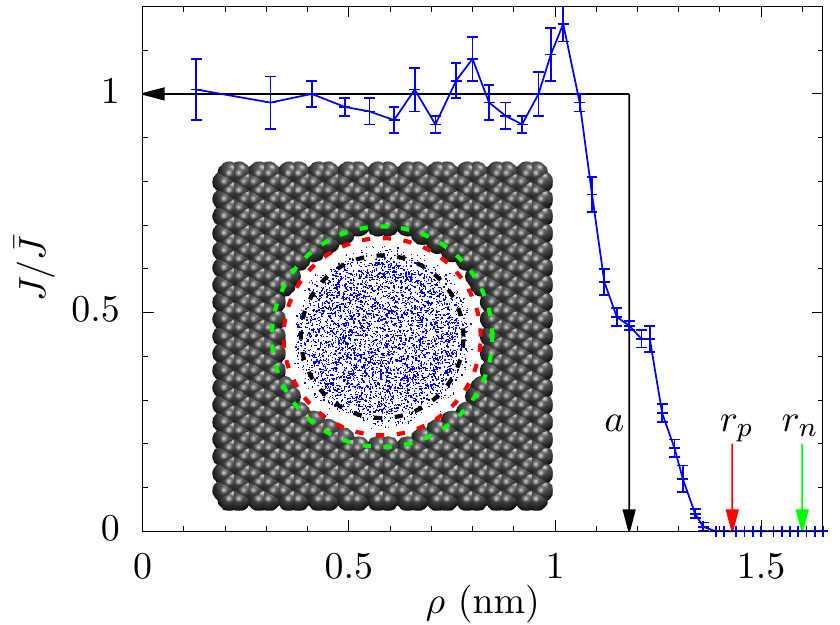}
\caption{Current density $J$ normalized with respect to its flat region. The effective pore radius $a=1.08$ nm is shown by the vertical black arrow; a pore with radius $a$ and a uniform current density $\bar{J}$ gives the same total current as the exact distribution, $\pi a^2 \bar{J} = I$.  The green arrow shows the largest circle going to the atom locations ($r_n$) and the red arrow shows the largest circle going to atom locations minus the vdW radius of carbon ($r_p$). The inset shows the structure of the pore in the vdW representation and the scatter plot of ions crossing the pore~\cite{sahu2018maxwell}. \label{fig:radius}}
\end{figure}

The effective radius should also include the finite size of ions (with hydration) and fluctuations of the pore edge \cite{sahu2018maxwell}. This can be given a rigorous form by calculating an ``unattenuated'' current density profile through the pore, and using this to set the effective pore edge; see Fig.~\ref{fig:radius}. This ignores geometric imperfections and graphene structure -- it considers all pores as circular. Further investigation is needed to know how these factors affect the access and pore resistance. Nevertheless, it is clear that {\em contextual} effects -- fluctuation of the pore edge, interaction between the edge and ions, and the size of the hydrated ions -- influence not only the conductance but the very definition of pore properties, such as radius and length. Their inclusion requires an in-depth analyses of all-atom MD results.

The discrepancies present in both computational and experimental results are thus resolved. Under reasonable conditions ($\lesssim 3$~M ion concentrations, $\lesssim 1$~V bias for computations and $\lesssim 0.25$~V for experiments), uncharged monolayer graphene has an effective thickness of about 1 nm. Pore radii above this value start to have a dominant access contribution, giving a resistance that scales inversely with radius. Pore radii around this value have contributions from both pore and access components to the resistance. Pore radii below this value we will address in Sec.~\ref{sec:Dehydration}, as dehydration comes into play.

\begin{figure*}
\includegraphics{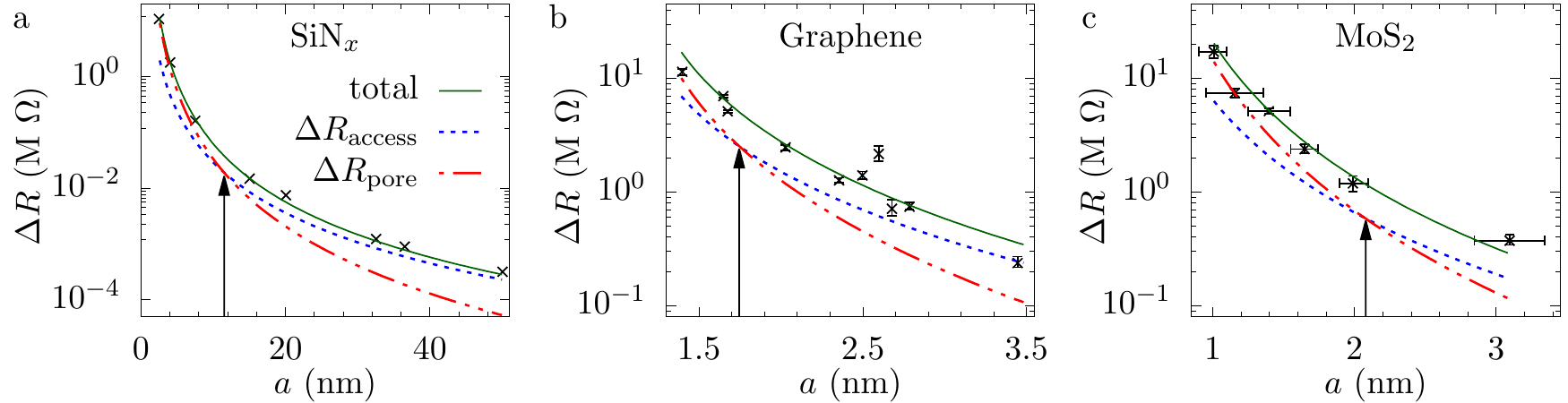}
\caption{Change in resistance, $\Delta R$, versus pore radius, $a$, due to current blockade by (a) dsDNA in SiN$_x$~\cite{Kowalczyk2011}, (b) dsDNA in graphene~\cite{garaj2013}, and (c) single A nucleotide in MoS$_2$~\cite{feng2015identification}. The open pore resistance is taken as $R = R_\access+R_\pore$, where $R_\access=\gamma/2a$ and  $R_\pore=\gamma h_p^{eff} /\pi a^2$. The blockade resistance is thus $\Delta R =  \Delta R_\access+\Delta R_\pore$, where $\Delta R_\access $ and  $\Delta R_\pore$ are changes in their respective resistances due to a change in pore radius to $a'=\sqrt{a^2-a_\mathrm{DNA}^2}$. We see that, for small pores, $\Delta R \approx \Delta R_\pore$ and, for large pores, $\Delta R \approx \Delta R_\access$. The transition from $\Delta R_\pore$ to $\Delta R_\access$  occurs when $a \gtrsim 4 h_p/\pi$ (assuming $a \gg a_\mathrm{DNA}$), as indicated by the arrows. The model works really well for SiN$_x$ (without fitting parameters) and reasonably for MoS$_2$, but only marginal for graphene (potentially due to sample-to-sample variation in pore structure/functionalization). For graphene and MoS$_2$, we use $\gamma$ as a fitting parameter due to the unknown local ion concentration during the blockade event. Error bars are shown when reported in the original article. \label{fig:dR}}.
\end{figure*}

\subsection{Implications}
The basic considerations of pore resistance versus access resistance has implications for sensing and simulations, and the interpretation of experimental data. In sensing, the blockade current -- how much a molecule or particle translocating through a pore blocks ion flow -- depends on being in a pore- or access-dominated regime. Using Eq.~\eqref{eq:TotalRes}, the change in resistance during a blockade event is the sum of access and pore contributions
\begin{equation}
\Delta R_\access + \Delta R_\pore=  \left( \frac{\gamma}{2 a^\p} -\frac{\gamma}{2 a} \right)  + \left( \frac{\gamma^\p \, h_p}{\pi {a^\p}^2} - \frac{\gamma \, h_p}{\pi a^2}  \right), \label{eq:dR}
\end{equation}
where $\gamma^\p$ and $a^\p$ are the resistivity and the radius of the pore in the presence of the translocating species. Determining $\gamma^\p$ is a challenge since it can increase or decrease depending on the pore charge, concentration of the electrolyte~\cite{smeets2006salt}, and other interactions. Nonetheless, at high salt concentration and relatively large pores, the effect of surface charge is small. 

Several works have shown that the geometric model in Eq.~\eqref{eq:dR}  (or some variation of it) explains the blockade current in nanopores. \citet{Kowalczyk2011} obtained a good fit to the experimental data for SiN$_x$ by taking $\gamma^\p=\gamma $ and $a^\p=\sqrt{a^2 -a^2_\mathrm{DNA}}$ with $a_\mathrm{DNA}=1.1$~nm for dsDNA; see Fig.~\ref{fig:dR}. This model fits the MoS$_2$ data from \citet{feng2015identification} fairly well but is marginal for graphene as observed by~\citet{garaj2013}. A similar model was also employed by \citet{wanunu2010rapid} to fit the blockade current due to DNA in SiN$_x$ pores. They took $\gamma^\p = 1/(n\,e\,[\mu_\Km + (1-S) \mu_\Clm])$, where $S$ is a fitting parameter to account for the reduced concentration of \Cl within the pore due to the presence of DNA. 

It is clear from Eq.~\eqref{eq:dR} and Fig.~\ref{fig:dR} that for smaller pores ($a<h_p$) the blockade current is influenced mainly by the change in pore resistance, whereas for larger pores ($a>h_p$) access resistance is more significant. This entails that the blockade current in 2D membranes is determined by the change in pore resistance only when the pore radius is below one to two nanometers. Once the pore radius exceeds this range access resistance plays the major role.

Working in the pore dominated regime seems desirable as the resistance change is largest there; see Fig.~\ref{fig:dR}. However, the resistance change in isolation is not what one wants to increase. Rather, one wants the highest signal-to-noise ratio (SNR). An intuitive account of the SNR is as follows. When $a$ is significantly larger than the DNA radius $a_{\mathrm{DNA}}$, the change in the pore radius for a blockade event is $\delta \approx - a_{\mathrm{DNA}}^2/2a$. In either the pore dominated or the access-dominated regime, the change in the current due to the blockade is inversely proportional to the total open pore resistance, i.e., $\Delta I \approx  (V/R) 2 \delta/a$ (pore dominated) or $\Delta I \approx  (V/R) \delta/a $ (access dominated). The noise depends on frequency: at low frequency there is a large $1/f$ noise and  at high frequency there is a large capacitive noise~\cite{wanunu2012nanopores, smeets2008noise}. At intermediate frequencies, the noise is given by thermal Johnson noise  $I_{rms}=\sqrt{4 k_B T \Delta f/R}$ ($k_B$ is the Boltzmann constant, $T$ is the temperature, and $\Delta f$ is the measurement bandwidth). Considering only the latter for simplicity, then  
\begin{equation} \label{eq:SN}
\mathrm{SNR}=\frac{\Delta I}{I_{rms}} \propto \frac{\delta}{\sqrt{a^2R}}\propto \frac{a_{\mathrm{DNA}}^2}{a \sqrt{a+2 h_p/\pi}} .
\end{equation}
Thus, both the pore thickness and the radius should be small for the highest SNR. This is in qualitative agreement with the results of~\citet{wanunu2010rapid}, where the SNR for sensing microRNAs using SiN$_x$ increases with decreasing membrane thickness and decreasing pore diameter. Intuitively, SNR decreases with $a$ because ions flowing far from the DNA/RNA in the pore adds noise but do not contribute to the signal~\cite{comer2016dna}. Similarly, both signal $\Delta I \sim 1/h_p $ and noise $I_{rms}\sim 1/\sqrt{h_p}$ decrease with height but the signal decreases faster and so the SNR decreases with height. 

Besides a high SNR for individual blockade events, it is desirable for distinguishing bases to have a high sensitivity~\cite{garaj2013, comer2016dna} 
\begin{equation}
\mathcal{S}= -\frac{\partial \Delta I}{\partial a_\text{DNA}}= \frac{V  } { R^2 } \frac{\partial \Delta R}{\partial a_\text{DNA}} \approx \frac{2\, \gamma^\p \,a_\text{DNA}\, V\, h_p}{\pi\, {a^\p}^4 R^2}  ,
\end{equation}
where we used $\Delta R$ from Eq.~\ref{eq:dR}, and assumed $a^\p=\sqrt{a^2 -a^2_\mathrm{DNA}}$ to be small, clearly necessary for large sensitivity. Initially, $\mathcal{S}$ can increase with $h_p$~\cite{comer2016dna}. However, as $h_p$ gets larger, $\mathcal{S}$ will decrease due to an increase in $R$. Sampling of multiple bases when $h_p$ increases beyond $\approx 1$ nm will further reduce distinguishability in the context of sequencing.

Additionally, there is another requirement for discriminating DNA bases: The pore should be small to ensure that the strand goes through one region at a time and not angled or entangled. These implications are in line with results for MoS$_2$ nanopores where different single nucleotides could not be distinguished unless the pore radius was less than 1.8 nm (about the same size as the effective membrane thickness 1.6 nm)~\cite{feng2015identification}. The above does not include the effect of dehydration, the counter-ion cloud, and DNA-pore interactions, as well as other effects, which likely will result in a true optimum (i.e., really small radii will result in exponentially suppressed currents and clogging). An interesting open issue is to determine this optimum under less stringent assumptions, such as including the full noise spectrum and dehydration (that may already be important in  the MoS$_2$ results due to the small pore radius).

In addition to implications for sensing, the {\em simulation} of 2D membranes requires incorporating access resistance. This is emphasized by the discussion above: Pores that are optimum for sensing are likely to come in a regime where both pore and access resistance contribute to the current and blockade. A direct argument makes this obvious: For detection of nucleotides in ssDNA, the effective nucleotide radius is about 0.7 nm (ignoring base flexibility), and the pore must have a radius at or above this level. Whether graphene or MoS$_2$, this gives a regime where $h^{\mathrm{eff}}_p/a \approx 1$. If the simulation does not capture both, it will not give an accurate picture -- potentially not even qualitatively -- of the blockade levels and distinguishability of molecules/bases/etc. 

Prior to the scaling ansatz, no approach existed to capture both access and pore resistance in simulations, other than making the simulation cell large enough that corrections are small, which is generally prohibitive. Scaling requires multiple simulations with different cell sizes, which increases the computational cost but by less than an order of magnitude. It also suggests an interesting possibility that requires no extra computational resources and may even reduce them: if the aspect ratio of the simulation cell is chosen at some special value ($\gar$) -- the {\em golden aspect ratio}~\cite{sahu2018maxwell,sahu2018golden} -- then there will be zero finite-size corrections and one will obtain the infinite, balanced size result for a finite, small simulation.

It turns out that this golden aspect ratio exists, with values given just after Eq.~\eqref{eq:golden}, as demonstrated by continuum and MD simulations~\cite{sahu2018golden}. We expect that even in the presence of contextual properties -- pore charges, structural transitions, fluctuations -- the golden aspect ratio should exist and take on the same numerical value as in the continuum case. This is because all-atom MD, with sufficiently large system sizes and weak field gradients, approaches the continuum limit. Moreover, local disturbances decay away from the pore and, beyond a certain length scale, the scaling should be analogous to the uncharged, non-contextual case. 

The scaling approach and the golden aspect ratio ``completes the circle'' -- or, should we say, ``complete the spheroidal shell'' -- to setting up rigorous all-atom MD, Brownian, and continuum simulations for ion transport and comparing directly to experiment. Of course, one needs applicable and accurate force fields, sufficiently long simulations (e.g., to obtain a statistically significant number of ion crossing events), and uncertainty quantification. The approach has already resolved issues with graphene pores, including showing that the pore radius is indeed contextual and that this has to be accounted for when defining the accessible area for transport. The simulation technique will further allow a quantitative study of the influence of contextual properties -- geometric imperfections, the presence of charges/dipoles, and structural fluctuations -- on access resistance, including in other solid-state membranes and biological ion channels.


\section{Many-body ion transport}\label{sec:Interactions}

In the previous section, we considered transport in the 2D membrane as a continuum geometric obstruction. It is clear that in actual pores this simplistic view is not enough. Membrane or pore charges, local free-energy barriers (e.g., due to dehydration), and structural fluctuations/transitions can all introduce additional complexities into ion transport. In fact, we have already seen how some of these factors can be incorporated into effective, contextual geometric parameters which are essential when dealing with pores at the nanoscale. These effects, though, are more than just nuisances to be approximated away but rather an integral part of the process.

The so-called Poisson-Nernst-Planck (PNP) equations~\cite{chen1993charges,eisenberg1996computing} retain the continuum description of transport, but also allow for aspects of many of these factors to be incorporated. In this approach, Poisson's equation, 
\begin{equation}\label{eq:Poisson}
\nabla^2 \phi = \sum_\nu \frac{q_\nu n_\nu}{\epsilon },
\end{equation}
and the stationary Nernst-Planck equation, 
\begin{equation}\label{eq:NernstPlanck}
{\bf J_\nu} = -q_\nu \left(D_\nu \nabla n_\nu +  \mu_\nu n_\nu \nabla \phi \right),
\end{equation}
are simultaneously solved. Here, $\phi$ is the potential, $\epsilon$ is the permittivity of the medium, and $J_\nu$, $q_\nu$, $n_\nu$, $D_\nu$, and $\mu_\nu$ are the current density, charge, concentration, diffusivity, and mobility of the ion species $\nu$, respectively. These equations give the current density due to both drift and diffusion of charge carriers in an inhomogeneous medium (e.g., with surface charges and screening). 

Dehydration and coordination with specific functional groups (or inhomogeneities) require going further still. These effects demand the atomistic description provided by all-atom molecular dynamics simulations to get estimates of free energies and local potential profiles that can be incorporated into Eq.~\eqref{eq:NernstPlanck}. In what follows, we will bridge these two descriptions using aspects of each to highlight important phenomena in 2D membranes. We start with a general description of selectivity and then discuss specifics of dehydration and interactions. 

\subsection{Selectivity}
As the gatekeepers of the cell, biological ion channels show remarkable ability to selectively allow high flows of the certain species. Solid-state nanopores aim to replicate this for applications such as solvent recovery, dialysis, and desalination. Understanding the origin of selectivity is essential for engineering membranes for applications. Selectivity generally arises because different ion species interact with the pore differently, an intentionally vague statement indicating that this process is complex. We now delineate the important factors. 

Selectivity is most often quantified by measuring the membrane (or reversal) potential $E_m$ due to a concentration gradient across the membrane (although in some  cases, directly measuring the partial currents from different species is possible). Since one ion preferentially transports through the pore (or membrane itself), the electronic potential will increase on one side of the membrane (into a quasi-stationary state regime before the unpreferred ions rectify the electrostatic imbalance). The selectivity (measured as the permeability ratio) is then found indirectly via the Goldman-Hodgkin-Katz voltage equation~\cite{goldman1943potential}
\begin{align}\label{eq:GHK}
E_m & =  \frac{\kt}{e} \ln \left(\frac{\sum_{c} \mathcal{P}_{c} [c]_\mathrm{high} +\sum_{a} \mathcal{P}_{a} [a]_\mathrm{low}}{\sum_{c} \mathcal{P}_{c} [c]_\mathrm{low} +\sum_{a} \mathcal{P}_{a} [a]_\mathrm{high}}      \right),
\end{align}
where $\mathcal{P}_{c(a)}$ is the permeability of cation $c$ (anion $a$) and $[c(a)]_s$ is the cation (anion) concentration on the $s=$ high, low concentration side. The expression assumes that the permeabilities are constant in the pore and interactions between ions can be ignored~\cite{hille1999ion}. 

In the simplest case, differing mobilities can give an apparent selectivity. This is not selectivity in the usual sense, as even large pores can give such selectivity due to differing bulk mobilities; and this will typically be very weak. In nanoscale pores/channels, the mobility of ions can also be influenced by interactions with the surface and dehydration. For instance, the mobility decreases in solid-state nanopores~\cite{ho2005electrolytic} and 2D channels~\cite{esfandiar2017size}. The hydration state of the ion also matters, where certain hydration states can increase mobility due to metastability of water orientation~\cite{peng2018effect}. \citet{bhattacharya2011} estimated mobility in $\alpha$-hemolysin using MD via $\mu_\pore/\mu_\bulk = v_\pore/v_\bulk$, where $v$ is the velocity of the ion under a constant force. The result was a $\approx 2$ to 3 fold decrease in mobility, which is not surprising since ions interact with charged groups on the pore interior. This kind of mobility change can result in ``true'' selectivity, although still weak. For pores in 2D membranes, we expect that a change in mobility inside the pore will be less significant due to the short pore length and may not even be possible to define, although there have been some attempts~\cite{suk2014ion, feng2016single}.

In most cases, selectivity arises due to other factors. Membrane and pore charge give counterion over coion selectivity. For nano- and subnano-scale pores, selectivity can be merely due to size: different (hydrated) ions and molecules are simply sterically hindered from going through the pore or otherwise see a different effective pore area. These processes of partial or full exclusion have to account for the membrane/pore edge flexibility and the fact that hydration layers are not rigid (but can deform without substantial penalty so long as water molecules are not lost). Other than the relative impact of these factors in 2D membranes, size-based exclusion is not that different in 2D membranes versus longer pores.

Steric hindrance is really just the extreme limit of selectivity due to different free-energy barriers of certain species. Most pores of interest in biology and analysis require that ions dehydrate at least partially to translocate. This gives an energy barrier, one that can be offset by interaction with charged functional groups. 2D membranes can be quite different: Ions can maintain a substantial number of water molecules on either side of the pore when it is atomically thick, thus lowering the dehydration barrier. When the pore diameter reaches about 1 nm, these effects (dehydration and interactions) will become very important, and the picture of resistance in Sec.~\ref{sec:IonTransport} fails. Moreover, for charged membranes and pores, interactions can already be significant even for larger pores.

\subsection{Dehydration\label{sec:Dehydration}}
The strong electric field around dissolved ions forces the nearby water molecules to orient into hydration layers (or solvation shells). Hydration of ions is an important component of reactions in aqueous solution~\cite{ohtaki1993structure}, ion channels~\cite{Doyle98-1, zhou2001chemistry,noskov2007importance, corry2012mechanism, kopec2018direct}, and nanopore sequencing~\cite{bhattacharya2016water}. The first hydration layer is strongly bound to the ion -- its energy range from about 1 eV in monovalent ions to about 10 eV in bivalent ions~\cite{Zwolak10} -- and tends to move along with it; whereas, the second layer is only partially oriented~\cite{impey1983hydration}. The third hydration shell is diffuse and only weakly defined; bulk behavior starts to appear in this region. The water molecules that are tightly bound around ions in solution are sterically hindered from accompanying the ion in subnanoscale pores. Thus, some water molecules have to break off when ions pass through the pore and their removal results in a rearrangement of the other water molecules or functional groups. Shedding of water creates a dehydration barrier for ions to translocate through the pore.

A simple estimate of the free-energy barrier is
\begin{equation}\label{dF}
\Delta F_\nu = \eta\sum_i f_{i\nu} E_{i\nu},
\end{equation}
where $\eta\approx 1/2$ accounts for nonlinear effects~\cite{sahu2017dehydration}, $f_{i\nu}$ is the fractional dehydration, and $E_{i\nu}$ is the solvation energy of the $i^\mathrm{th}$ hydration layer in bulk~\cite{Zwolak09,Zwolak10} (we ignore entropic factors). The fractional dehydration depends on geometry and dimensions, which determine the volume available for the water molecules to solvate ions. It is given by 
\begin{equation}
f_{i\nu} = \frac{\Delta n_i}{n_i} \approx \frac{\Delta \mathcal{V}_i}{\mathcal{V}_i} ,
\end{equation}
where $n_i$ ($\mathcal{V}_i$) is the coordination number (volume) of the $i^\mathrm{th}$ layer in bulk, and $\Delta n_i$ ($\Delta \mathcal{V}_i$) is its change in the pore.

\begin{figure}
\includegraphics{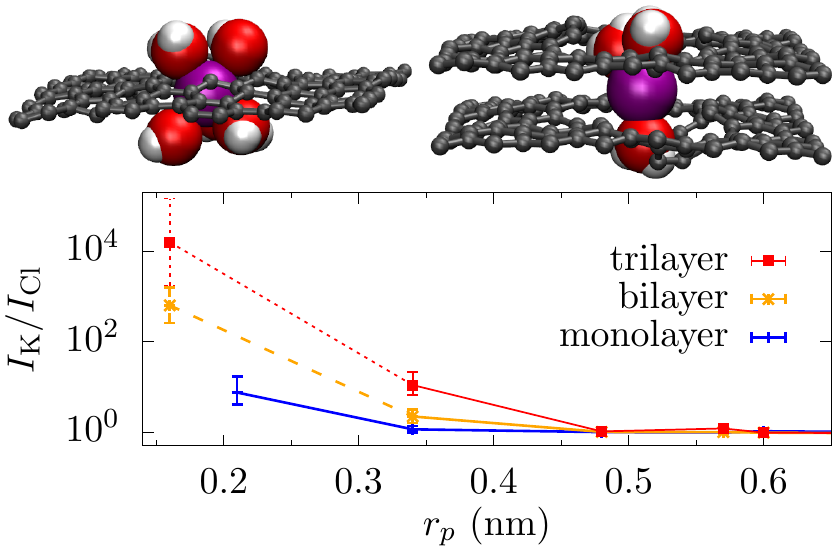}
\caption{\K (purple) translocating through mono- (left) and bi-layer (right) graphene pores (geometric radii of 0.2 nm and 0.16 nm, respectively). The carbon atoms are shown as smaller gray spheres (not the vdW radii like the other atoms) along with the carbon-carbon bond. For pores of this size, ions cannot retain the complete hydration shell when translocating. For monolayer graphene, \K loses roughly two water molecules from its first hydration shell but still retains four closely bound water molecules just outside the membrane [large red (O) and small white (H) spheres]. For bilayer graphene, however, water molecules can hydrate only on the ``two ends'' of the ion, which gives a substantially larger energy barrier. Tri-layer graphene further limits hydration. The bottom panel shows the \K over \Cl selectivity (given by the ratio of their currents, $I_\mathrm{K}/I_\mathrm{Cl}$) in graphene pores versus the geometric radius. The multilayer graphene is AB stacked, which influences the allowed radii. All data points are from nonequilibrium MD simulations~\cite{sahu2017dehydration,sahu2017ionic} except for the smallest pore in bi- (dashed line) and tri-layer graphene (dotted line), which were estimated from free-energy barriers. Lines are a guide to the eye only.\label{fig:dehydration}}
\end{figure}

While an all-atom description is necessary to get a quantitative account of dehydration (force field validity notwithstanding), the basic influence on resistance can be incorporated into a continuum description via a spatially dependent free-energy barrier. Solving the PNP equations~\eqref{eq:Poisson} and \eqref{eq:NernstPlanck} in one dimension -- a rather drastic approximation but one that captures the main features of transport -- the current density through the pore (at $z=0$) is~\cite{eisenberg1995diffusion}
\begin{equation}
J_\nu =  \mu_\nu k_B T \frac{n_+ e^{q_\nu \phi(h_p/2)/k_BT} - n_- e^{q_\nu \phi(-h_p/2)/k_B T}}{\int\limits^{h_p/2}_{-h_p/2}e^{q_\nu \phi(z)/k_BT} dz},
\end{equation}
where $n_\pm$ is the ion density at $z=\pm h_p/2$. Using a simple model for the potential $\phi(z)=z\,V_p/h_p+\Delta F_\nu/q_\nu$ with constant free-energy barrier $\Delta F_\nu$ along the pore length $h_p$ and taking $n_+=n_-=n_\nu$, the current is 
\begin{equation}\label{eq:current}
I_\nu  =  q_\nu \mu_\nu E A_p n_\nu e^{-\Delta F_\nu / k_B T},
\end{equation}
where $A_p$ is the pore area and $E=V/h_p$ is the electric field across the pore. Dehydration thus exponentially increases the resistance by, essentially, depressing the ion density in the pore as $n_\nu e^{-\Delta F_\nu / k_B T}$. We note that Eq.~\eqref{dF} is a phenomenological model of dehydration. The factor $\eta$ accounts for the stronger orientation of the remaining water dipoles~\cite{sahu2017dehydration}. Other many-body interactions, such as ion-ion interactions and polarization of water molecules by the applied field, also shape the free-energy landscape in the pore, as do various nonequilibrium factors.

\begin{figure*}
\includegraphics{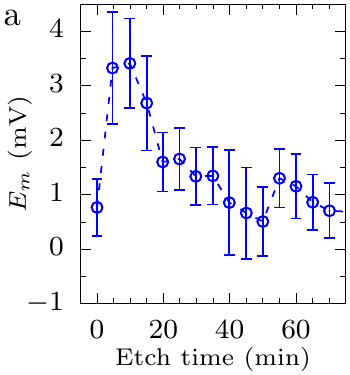}
\includegraphics{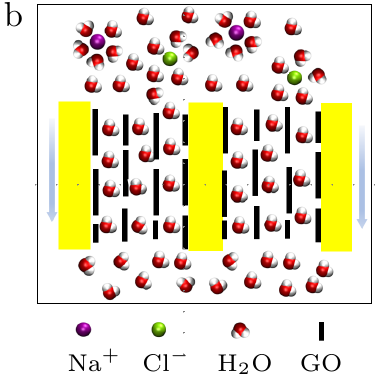}
\includegraphics{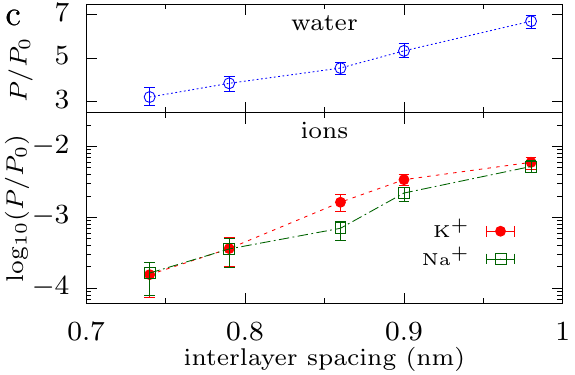}
\includegraphics{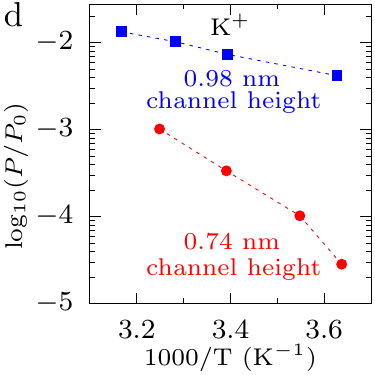}
\caption{Observations of dehydration-based selectivity in 2D membranes. (a) Membrane potential versus etch time (pore radius) showing weak selectivity in subnanoscale graphene pores~\cite{OHern2014}, consistent with dehydration. (b) Schematic of permeation through GO layers in the experiments by~\citet{abraham2017}. (c) Permeation rate, $P$, in units of $P_0= 1$ mol h$^{-1}$m$^{-2}$, of water and ions for the variable interlayer separation in (b). For water, the permeation increases linearly with increasing interlayer spacing, whereas for ions it increases exponentially. (d) Permeation rate for \K ions in (b) versus temperature showing Arrhenius behavior. All dashed connecting lines are guides to the eye only.\label{fig:selectivity}}
\end{figure*}

In long pores, dehydration should give a series of drops in the conductance versus pore radius as each hydration layer is partially excluded from the pore~\cite{Zwolak09,Zwolak10}, a phenomenon that is reminiscent of quantized steps in the conductance in solid-state systems. As the pore radius approaches the size of the first hydration layer, the dehydration barrier becomes prohibitively large ($> 0.5$~eV) \cite{beckstein2004not,Zwolak09,Zwolak10}. Unless there are charged groups to compensate, as in biological pores, ions cannot translocate under normal conditions. This makes it challenging to quantify dehydration alone: either the currents are too small to be measurable or, other interactions (such as electrostatics), obscure the effects of dehydration~\cite{noskov2006ion}. Compounding this difficulty is the fact that one needs atomic level control to make the pore radius comparable to the radius of hydration shells ($\approx 0.3$ nm to $\approx 0.6$ nm, ignoring the third layer, which however can be important for divalent ions~\cite{Zwolak10}). 

The large free-energy barrier is due to the fact that $\Delta \mathcal{V}$ is large when constricting in multiple directions. For very small radii ($r_p<0.5$ nm) and long pores ($h_p>2$ nm), only single water can hydrate an ion on each side, meaning that about 70 \% of all water molecules are blocked from the first solvation, and nearly all from the second and third solvation shells. This means that the hydration barrier is substantial, about 1~eV to 2~eV, when considering the magnitude of $E_i$ in Eq.~\ref{dF}. Pores in 2D membranes, while nominally constricting in two directions (the plane of the membrane), allow for ions to maintain substantial hydration on the two sides of the membrane; see Fig.~\ref{fig:dehydration}. This phenomenon is unique to single atom thick membranes. Already for very small radii pores in bilayer graphene that encroach on the first hydration shell, ions can maintain only one or two water molecule on either side, leading to substantially larger free-energy barriers~\cite{sahu2017ionic}. The generally smaller barriers and tunability of both radius and number of layers should enable the measurement of dehydration; and, when functional groups are present, its competition with local interactions. 

Since the dehydration energy influences the permeation of ion through the pore, it also leads to ion selectivity~\cite{kopec2018direct, song2009, Zwolak09}. The smaller barriers for 2D membranes should allow for the direct measurement of dehydration-only selectivity~\cite{sahu2017dehydration}; which may also be possible with particular carbon nanotubes~\cite{song2009}. For example, due to a smaller dehydration energy, \K is selected over \Cl by neutral subnanoscale graphene pores despite both ions having similar hydrated sizes and mobility~\cite{sahu2017dehydration}. The lower panels in Fig.~\ref{fig:dehydration} show the selectivity of \K over \Cl in mono-, bi-, and tri-layer graphene membranes. For monolayer graphene, a geometric radius of about 0.2 nm gives a selectivity factor of 3 to 10, and measurable currents of 32 pA to 48 pA and  6 pA to 9 pA (depending on water model used~\cite{sahu2017dehydration,sahu2017ionic}).

Bi- and tri-layer graphene are already selective at a pore radius of about 0.35 nm (due in part to second hydration layer exclusion). For tri-layer pores, this selectivity is about a factor of 12 and also measurable with currents of 59 pA and 5 pA. Layering in 2D membranes, thus, may even give rise to quantized selectivity. The major open issue is that the pore radius itself is ``discretized'' at the atomic scale. The spacing between points in Fig.~\ref{fig:dehydration} is reflective of this; these are all the symmetric pores allowed in graphene. To get intermediate radii, one would have to have nonsymmetric pores and ``radius'' would only be a crude measure of size (and dehydration can be very sensitive to the precise geometry). Thus, the very notion of sharpness is unclear for radii at the atomic scale.

In fact, this dehydration-only selectivity may have already been seen in the monolayer experiments of \citet{OHern2014}. They found a membrane potential of 3.3 mV (Fig.~\ref{fig:selectivity}a) for \K and \Cl for a very small average pore radius (the fabrication technique was designed to make many pores of more or less the same size, see Sec. II). Estimating the selectivity from Eq.~\eqref{eq:GHK} and the distribution of pore sizes gives a selectivity factor of about 2~\cite{sahu2017dehydration}. This is in line with the magnitude of selectivity due to dehydration alone. However, \citeauthor{OHern2014} attributed it to negatively charged functional groups terminating the edge of the pore. One can not rule out a charged-based mechanism from the existing data. Nonetheless, charge tends to give much larger selectivities and also can persist for larger pore sizes. There has also been a report of dehydration-based selectivity among cations in graphene pores~\cite{jain2015heterogeneous}, but this result is consistent with just differing bulk mobilities~\cite{sahu2017dehydration} and could also be a charged-based selectivity among cations, which is not generally strong, as seen by~\citet{rollings2016}.

Channels made from 2D heterostructures or graphene oxide (GO) laminates also give less confinement than long pores and provide a platform for observing dehydration-based selectivity. In these systems, there is now direct evidence of dehydration dominated selectivity and even results quantifying the dehydration barrier (albeit with some contribution from interaction with oxide functional groups) and supporting the notion of quantized conductance proposed by~\citet{Zwolak09}. \citet{Joshi2014}, for instance, examined permeation across GO laminates, which have two-dimensional channels due to hydration of the space between GO layers. The layer spacing was not adjustable; however, hydrated ions larger than the layer spacing (about 0.9 nm) were prevented from permeating, giving a sudden drop in conductance.

Later, the same group invented a technique to change the layer spacing~\cite{abraham2017}. They first swelled the GO laminate in vapor conditions to limit the amount of hydration. Before placing in water (which would swell it further), they encapsulated the laminate in an epoxy, which fixed the layer spacing (Fig.~\ref{fig:selectivity}b). This allowed them to examine the permeation rate versus layer spacing; see Fig.~\ref{fig:selectivity}c. This dependency is analogous to an Arrhenius plot (rate versus temperature, see Fig.~\ref{fig:selectivity}d) to extract the free-energy barrier, but here one can obtain the free-energy barrier dependence on radius. It shows that $\Delta F$ decreases linearly with the layer spacing so long as the channel height (minus extra spacing due to functional groups) is encroaching on the hydration layers. This supports the excluded volume model in Eq.~\eqref{dF} when applied to the channel geometry. 

This barrier, though, still includes interactions with the oxide functional groups. Regardless, this is a pioneering experiment on a fundamental aspect of ion transport at the nanoscale. In a new experiment~\cite{esfandiar2017size} from the same lab, the effect of hydration layers is delineated more precisely by using graphene and MoS$_2$ as spacers to create channels in stacks of hBN, graphite, or MoS$_2$. These results showed a clear relation between channel conductivity and size of the hydration shell. This lab has taken this approach a step further, showing that single-layer high channels let only protons through~\cite{gopinadhan2019complete}. Similar experiments for porous systems will shed light on the complex array of processes occurring in biological systems. All in all, the ``phase space'' of experiments, between 2D heterostructures, GO membranes, pores in 2D membranes, will bring about a rigorous treatment of dehydration and interaction in nanoscale ionic transport.

\subsection{Fixed charges}\label{sec:charges}
Surfaces generally carry fixed charges which attract counterions and repel coions from the surrounding solution, thus forming the well-known electrical double layer (EDL). The important length scale for this effect is the Debye length, $\lambda_D= \sqrt{\epsilon \kt/\sum_\nu n_\nu q_\nu^2 }$, which is the distance at which the  surface charge is effectively screened by the ions in solution. When the pore radius is comparable to or smaller than the Debye length, the EDLs from opposite sides of the pore interior overlap and the coions can be completely excluded. The effect of surface charge is most pronounced in such a case.

\begin{figure}
\includegraphics{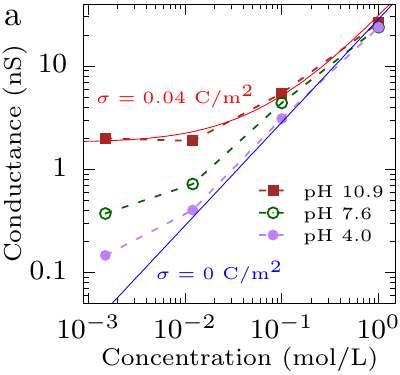}
\includegraphics{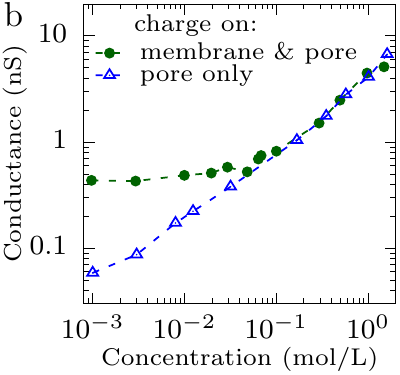}
\caption{Saturation of the ionic current due to membrane charge. (a) Conductance versus ion concentration in a stacked graphene-Al$_2$O$_3$ pore of diameter 8 nm and length $h_p=20$ nm, where the surface charge is controlled by varying the \pH of the solution~\cite{venkatesan2011stacked}. The two continuous lines show the  conductance with no surface charge and with surface charge obtained by fitting data for \pH 10.9 to Eq.~\eqref{eq:G2}. (b) Conductance versus ion concentration in the bacterial porin OmpF~\cite{alcaraz2017ion}. The membrane surface needs to be charged (green circles) for the current to saturate, whereas for pore charge only (blue triangles) it does not saturate. Dashed lines are guides to the eye only.\label{fig:saturation}}
\end{figure}

When the counterions on a charged surface flow due to an applied field, the induced surface current is
\begin{equation}\label{eq:surfaceI}
I_\surface=  2\pi a \sigma \mu  E =k_s 2\pi a E,   
\end{equation}
where $\sigma$ is the surface charge density, $\mu$ the mobility of the counterion, and $k_s=\sigma \mu$ is the surface conductivity. This is in addition to the usual volume current,
\begin{equation}
I_\volume= \sum_\nu  e \pi a^2 \mu_\nu   n_\nu E = k_b \pi a^2 E,
\end{equation}
where $k_b=\sum_\nu n_\nu \mu_\nu e$ is the bulk conductivity (we have ignored complications from pore/surface dependent mobilities, etc.). The surface-to-volume current ratio is then
\begin{equation} \label{eq:CurrentRatio}
\frac{I_\surface}{I_\volume} = \frac{ 2k_s}{ a k_b}  = \frac{2 \LD}{a }
\end{equation}
where $\LD=k_s/k_b\approx \sigma/2e n_\bulk$ is the ``Dukhin length"~\cite{bocquet2010nanofluidics}.  Equation~\eqref{eq:CurrentRatio} shows that at low concentrations (and high surface charge density), the surface current dominates over the volume current. Since the surface current is independent of the bulk concentration, it should saturate at low ion concentration~\cite{schoch2005ion, stein2004surface, feng2016single, weber2017boron,alcaraz2017ion}. 

Current saturation in stacked graphene-Al$_2$O$_3$ pores is shown in Fig.~\ref{fig:saturation}a~\cite{venkatesan2011stacked}. As with all surfaces, the magnitude of the effective surface charge depends on the \pH of the solution~\cite{parks1965isoelectric}. As seen in these experiments, the lower the \pH, the smaller the surface charge. Saturation, therefore, occurs at smaller and smaller concentrations. Biological systems also display these effects; see Fig.~\ref{fig:saturation}b. 

Even though the pore conductance saturates at low concentration, the current may not saturate due to access resistance~\cite{song1999meningococcal, alcaraz2017ion}. Often, in larger solid-state channels, the conductance is measured with the electrodes close to the channel entrance/exit, in which case only the pore conductance is measured. However, when the electrodes are far away, we have to consider two cases: (1) both the pore and membrane surfaces are charged and (2) only the pore is charged.  \citet{lee2012large} showed that when both the pore and membrane surfaces are charged, the effective pore diameter for access resistance increases from $2a$ to $2a+\LD$. In addition, the pore resistance decreases due to the surface current. Thus the total conductance is
\begin{equation}\label{eq:G1}
G_1 =k_b \left( \frac{1}{2a + \LD } + \frac{h_p}{\pi a^2 + 2\pi  a  \LD} \right)^{-1} .
\end{equation}
At low bulk concentration, i.e., $\LD\gg a$, the conductance saturates to 
\begin{equation}
G_1(\LD\gg a) =k_s \left( 1 + \frac{h_p}{2\pi  a} \right)^{-1} .
\end{equation}
Thus, the charged membrane substantially reduces the access resistance as seen in \citet{aguilella2005}. This causes the saturation of current at low concentration (i.e., when the surface current dominates).

When only the pore is charged, the effective pore radius for access resistance increases by $\lambda_D$ \cite{peskoff1988electrodiffusion} and  the pore resistance is reduced by the surface current giving the conductance
\begin{equation}\label{eq:G2}
G_2=k_b \left( \frac{\chi}{a+\lambda_D} + \frac{h_p}{\pi a^2 + 2\pi  a  \LD} \right)^{-1},
\end{equation}
where the factor $\chi$ depends on concentration~\cite{levadny1998} and pore charge~\cite{aguilella2005}. At low ion concentration, the conductance is 
\begin{equation}
G_2 (\LD \gg a) = k_s \left( \frac{\chi\LD}{\lambda_D} + \frac{h_p}{2\pi  a} \right)^{-1},
\end{equation}
which still depends on concentration. This expression has not been verified experimentally to our knowledge. In some biological settings, the pore and membrane charge can be set independently, going between these regimes; see Fig.~\ref{fig:saturation}b. An appropriate 2D membrane may also be able to tune these charges separately. Thus, there may be opportunities to quantify the influence of charge (and non-ideal geometries) on access resistance.

\begin{figure}[t]
\includegraphics{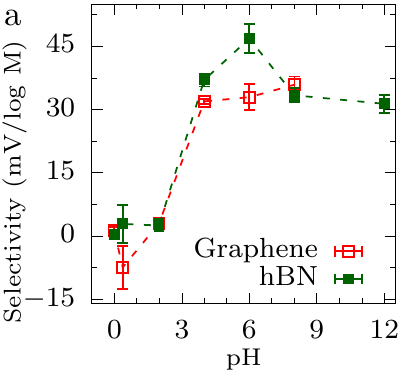}
\includegraphics{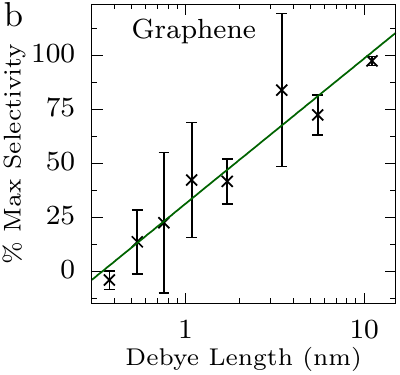}
\includegraphics{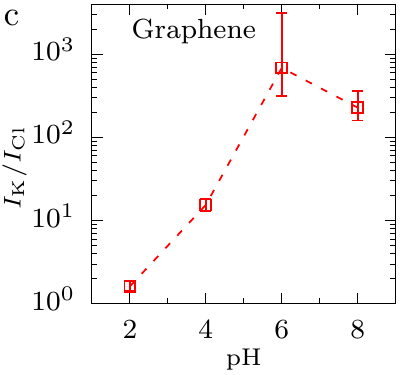}
\includegraphics{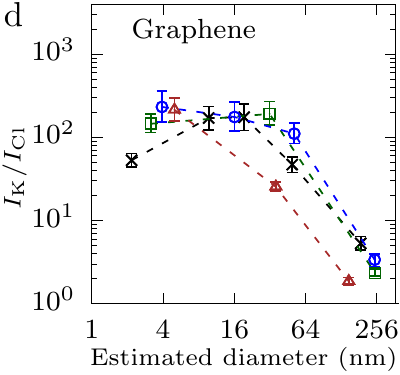}
\caption{Charged-based selectivity. (a) Selectivity increase with \pH (graphene and hBN) and (b) with Debye length (graphene)~\cite{walker2017}. (c) A sharp increase in \K selectivity with \pH for a graphene pore of diameter 3 nm and, (d) ion selectivity in graphene pores from several devices (shown with different markers)~\cite{rollings2016}. Dashed lines are guides to the eye only.\label{fig:X}}
\end{figure}

The attraction of counterions and repulsion of coions also makes the pore selective. This is often the case in biological systems, such as acetylcholine~\cite{unwin2005refined} and Cys-loop receptors~\cite{hibbs2011principles}. Selectivity due to membrane charge is also seen in 2D membranes. \citet{walker2017} observed cation selectivity via the reversal potential in graphene and hBN pores (estimated to be 0.4 nm to 3 nm in diameter) made by ozone treatment and chemical etching. The selectivity depends on both the Debye length and solution \pH, see Figs.~\ref{fig:X}a,b, indicating that it is due to charge.

The Debye length determines the spatial exclusion of ions with like charge to the surface. In long pores, charge-based selectivity is primarily determined by the low concentration side due to lack of electrostatic shielding of the surface on that end and the well-separated nature of the two sides. Unlike long pores, this selectivity in 2D membranes is controlled by the Debye length of the {\em high} concentration side of the membrane. \citet{walker2017} attribute this to ion exclusion at the pore mouth. Due to the small channel length, the high concentration side determines the exclusion near the pore on {\em both} sides to the membrane. A theoretical demonstration of this phenomenon, especially the associated ion densities and screenings, is an important open issue: The short channel length means that the two sides of the membrane can not be considered -- not even approximately -- as independent reservoirs at some given concentration. For instance, if the high concentration side partially diffuses through to the low concentration side, screening the pore opening locally, then one cannot easily interpret the selectivity measurement. In fact, it is already nontrivial to understand membrane/reversal potential measurements~\cite{alcaraz2004salting}. Due to the high concentration side influence, they may underestimate the selectivity in 2D membranes if interpreted as long pores. This is one factor that makes it difficult to determine if the weak selectivity seen in other experiments~\cite{OHern2014} is due to fixed charges or dehydration.

Since electrostatic interaction is a comparatively long-range, especially in low salt solutions, the selectivity due to charge will persist in pores larger than the ionic hydration radius -- at least to a couple Debye lengths~\cite{schoch2008transport}. Moreover, when the membrane is also charged, the selectivity depends on the surface-to-volume current ratio (i.e., the strength of the surface charge). Thus, charge-based selectivity with a charged membrane and pore can extend to very large pores, so long as the surface charge is substantial. \citet{rollings2016} found strong selectivity that persists for very large pores (past 20 nm in diameter); see Figs.~\ref{fig:X}c,d.

These results, however, do not tell us what is responsible for the surface charge. Is it, for instance, functional groups? Since both graphene and hBN -- chemically very different materials -- show the same behavior, \citet{walker2017} attribute the presence of surface charges to {\em extrinsic} factors, namely the adsorption of OH$^-$  groups. This is consistent with the selectivity seen by~\citet{rollings2016} in graphene and the disappearance of selectivity at low \pH, although this does not substantially narrow down the negatively charged groups present.

\subsection{Functionalization}
As discussed in Sec.~\ref{sec:Classes}, functional groups play a major role in gating, selectivity, and permeability of biological channels. Functional groups can introduce partial charges, but the effect is more than the continuum influence of the last section. In biological settings, selectivity is often based on the placement of charged functional groups: how they coordinate with ions to balance dehydration and how they change during gating and structural transitions. Researchers can employ targeted mutations on natural (wild type) ion channels to alter specific functional groups~\cite{heginbotham1994mutations, merzlyak2005conductance} to investigate the function of those sites. Controlled functionalization of graphene and other 2D materials will make it possible to mimic properties of selectivity and permeability of biological channels. 

Functionalization involves depositing or covalently binding active material in the pore and membrane surface which changes their physical and chemical properties. In graphene electronics, for example, often the purpose of functionalization is to open a band-gap or to change the surface chemistry~\cite{bellunato2016chemistry}. For ion transport, the variation of surface characteristics due to functionalization can substantially alter the transport properties of the pore. Functionalization of the membrane surface is also critical. For example, the surface coating can prevent DNA from sticking on a graphene surface or give antifouling properties (this is in addition to the introduction of surface charge). 

Several methods have been studied for controlled functionalization of graphene. In reactive plasma etching, the graphene substrate reacts and forms a compound with atoms in the plasma~\cite{bellunato2016chemistry}. Graphene can also be functionalized by exfoliating it via chemical reactions~\cite{economopoulos2010exfoliation}. Most of these techniques are developed for functionalizing the outer edge of a graphene ribbon, for example, with oxygen \cite{wang2010etching} or hydrogen~\cite{xie2010selective}. Selective functionalization of a pore edge will be more challenging. Besides chemistry, the steric hindrance and mechanical stress also play role in the functionalization of the pores~\cite{bellunato2016chemistry}.

Graphene is chemically stable due to its aromatic structure; however, edge atoms in the pore with unsaturated bond are chemically active~\cite{bellunato2016chemistry}. Thus, graphene pores immersed in an electrolyte or exposed to air will end up passivated. Several studies indicate that graphene nanopores can possess negative surface charge~\cite{shan2013surface}. The nature of functionalization of the carbon edges in the pore has not been pinpointed. \citet{rollings2016} suggest that the likely mechanism is the oxidation of carbon atoms at the edge, leading to the formation of carboxyl groups. \citet{OHern2014} also indicate that different oxygen-containing groups are passivating the pore edge and causing pore enlargement (via KOH etching) to halt. The most common functionalized graphene is the GO membrane, which has been studied mainly in the context of separation~\cite{nair2012, Joshi2014,abraham2017}. This membrane is functionalized everywhere, averaging transport properties as an ion goes through a channel. So far both control and characterization of only the pore edge has been elusive.

\begin{figure}
\includegraphics{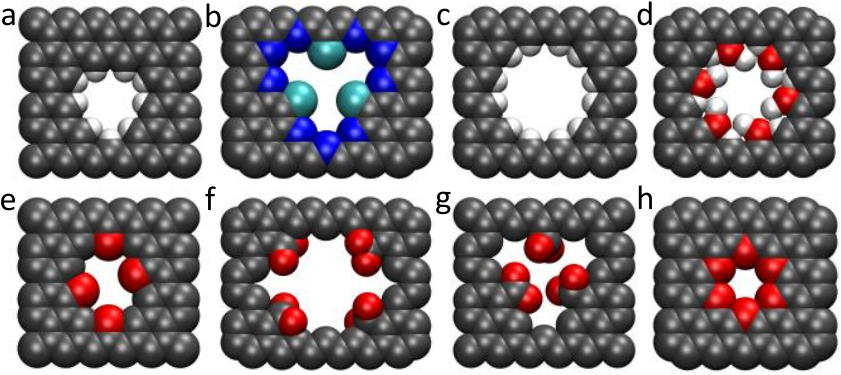}
\caption{Some examples of functionalized graphene pores: (a) hydrogen (white) terminated and (b) fluorine-nitrogen (blue-green) terminated graphene nanopores~\cite{sint2008}. (c) Hydrogenated and (d) hydroxylated graphene pores~\cite{tanugi2012}. Graphene nanopores functionalized with (e) four carbonyl, (f) four carboxylate, and (g) three carboxylate groups~\cite{he2013bioinspired}. (h) Crown-ether graphene~\cite{guo2014crown}. The pores in (a-g) are hypothetical but variants of (h) have been seen in experiment.\label{fig:functionalized}}
\end{figure}

In this regard, MD studies have been employed to elucidate the role of functionalization in pore properties. \citet{sint2008} studied ion transport through functionalized graphene pores; see Figs.~\ref{fig:functionalized}a,b. They showed that only cations can translocate through fluorine-nitrogen terminated pores, whereas only anions can translocate through hydrogen terminated pores. Additionally, the flow rate varied between cations and anion depending on the strength of their hydration shell. \citet{tanugi2012}, similarly, found that hydrogenation (Fig.~\ref{fig:functionalized}c) lowered water transport but hydroxylation Fig.~\ref{fig:functionalized}d) enhanced it. However, salt rejection was higher in hydrogenated pores compare to hydroxylated ones. Essentially, the hydroxylation lowers the barrier to transport, as the OH$^-$ functional group can form hydrogen-bonds with water dipoles and ions. \citet{he2013bioinspired} examined transport in functionalized graphene pores, shown in Fig.~\ref{fig:functionalized}e-g: A 4-carbonyl (4CO) pore that resembles KcsA and 4-carboxylate (4COO) pore that resembles NavAb. 4CO is \K selective over \Na like its biological counterpart, but the selectivity is orders of magnitude weaker.  In contrast to its biological counterpart, 4COO pore was \K selective rather than \Na selective. However, the 3-carboxylate (3COO) pore they studied may have strong \Na over \K selectivity at lower voltages, but this regime was not examined in detail.

An exciting example of a functionalized pore is the formation of crown ether structures in graphene~\cite{guo2014crown}. This pore has been observed in experiments and partially characterized via TEM (but ion transport has not been measured). Stand-alone crown ethers have been studied for a long time for their property of selectivity and ``host chemistry''. However, the flexibility of the stand-alone crown ether significantly reduces its selectivity and binding strength. In the symbiotic crown ether-graphene, the pore provides a selective binding site for cations in graphene, and the graphene gives the rigid plane structure to the crown ether. Since the type of crown ether determines the selectivity~\citep{guo2014crown}, different crown ether graphene can potentially be used as a sensor for different metal ions. Due to the structural similarity to the selectivity filter, crown ether and related pores will open up many opportunities to study the transport mechanisms in biological channels, as well as for filtration and desalination technologies.

\subsection{Implications}
Graphene and other 2D membranes are interesting because they offer a novel testbed to study permeation and selectivity from the ground up while yielding other opportunities for sensing and filtration. For the former, in particular, one can build a channel layer by layer, going from mono- to bi- to tri-layer graphene. Thus, both the pore radius and thickness can be tailored. Moreover, the carbon allotrope allows for chemical functionalization with important groups that can incorporate dipoles and charges into the pore interior. Even though not yet realized in a controllable way, this would be revolutionary in the study of biological and biomimetic channels.

While still on the horizon for functional groups and local interactions, 2D channels and graphene pores are already making inroads in this regard with dehydration and charged membranes, with other opportunities just around the corner (such as the effect of non-ideal geometries and interactions on dehydration). In addition, there are other many-body effects in nanoscale pores that cause nonlinearities in the current with respect to \pH, salt concentration, or voltage. One example is the result of ion-ion interactions. These are particularly strong in confined spaces, especially when hydration layers are broken (in bulk or large pores, the large dielectric constant of water, $\epsilon_r \approx 80$, gives a much weaker ion-ion interaction). Even without dehydration, the reduced dielectric constant in narrow pores and channels~\cite{fumagalli2018anomalously} due to the surface-induced alignment of water dipoles increases the ion-ion interaction. Thus, when there is a ``weak link'' for ions to enter and/or exit the pore, they can accumulate -- charging the ``pore capacitor'' -- and preventing others from going through, analogous to Coulomb blockade in quantum dots~\cite{krems2013, feng2016observation, tanaka2017surface}. Similar to its solid-state counterpart,  this so-called {\em ionic Coulomb blockade} will give nonlinearities in the current versus voltage, as only when the voltage is large enough to overcome the charging energy, will more current flow. 

This phenomenon may already have been seen in MoS$_2$ pores~\cite{feng2016observation}. There, in addition to a suppression of the current at low voltage, \citeauthor{feng2016observation} observe the oscillatory behavior of the differential conductance with respect to \pH. The \pH changes the surface charge and acts as a gate, giving Coulomb oscillations: When ions pay no energy penalty to get into or out of the pore, a higher current flows. However, whether unoccupied or when an ion is trapped in the pore (due to the local surface charge), there is an electrostatic penalty for an ion/another ion to come in. The gate lowers that penalty until an ion can come in, initially increasing the current, but further reduction results in the localization of the ion and the process repeats. This suggests many-body physics is at play. However, it is not clear one can rule out dehydration effects (the size of the pores are 0.6 nm to 0.8 nm): The interplay between partial dehydration and interaction with surface charges can shift the local free-energy minimum/minima, as well as alter the mobility (e.g., it can depend on the hydration state~\cite{peng2018effect}). Moreover, there is steric repulsion of ions. Within the relevant energy regimes, this might even give quantized effects -- when one, two, etc., waters peel off -- yielding an increasing ion density in discrete steps. Still, it is clear that the concepts of ionic Coulomb blockade, quantized conductance, and ``volume exclusion'' dehydration, Eq.~\eqref{dF}, are helping us to understand ion channels better ~\cite{fedorenko2018quantized, kaufman2015coulomb}, even though it has long been known that electrostatics, solvation, and functional groups play the  crucial roles. 

\begin{figure}
    \centering
    \includegraphics{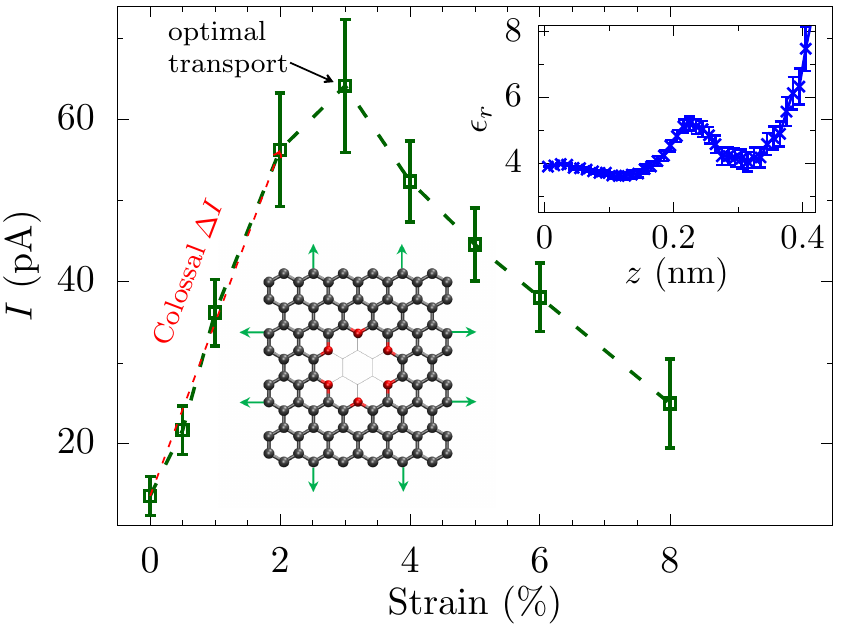}
    \caption{Colossal mechano-conductance and optimal transport in a graphene crown ether pore  (bottom inset)~\cite{sahu2019optimal}. Each oxygen and carbon at the pore rim has partial charge  $-0.24\, e$ and $0.12\, e$, respectively. The top inset shows the effective dielectric constant ($\epsilon_r$) near the pore center. Small changes in the pore size (i.e., 1 \% to 2 \%) due to strain result in a large (i.e., 200 \% to 300 \%) change in current. This is driven by a flattening of $\Delta F$ versus $z$ -- i.e., a tendency toward barrierless transport -- but ultimately the charged groups do not compensate for dehydration and a larger barrier decreases $I$.}
    \label{fig:I-strain}
\end{figure}

These concepts culminate in functionalized, 2D pores. For instance, \citet{sahu2019optimal} recently report on the effect of strain on the ionic current. A minuscule change in pore size, e.g., via strain, induces a colossal change in the ionic current through charged subnanoscale pores, such as graphene crown ethers and biological channels, see Fig~\ref{fig:I-strain}. While a few percent change in pore size does not change the dehydration barrier in some cases, it does significantly change the electrostatic energy because of the distances involved and the reduced dielectric screening in the pore. A mere 10 pm change in the radius of pore (initially 0.28 nm) shown in Fig~\ref{fig:I-strain} gives rise to a 1 $\kt$ change in a translocating ion's energy barrier which leads to about a three-fold change in the current. This gives an effective method for modulating ion transport toward a barrierless regime and optimizing currents. In addition, this modulation can also change the ion transport mechanism from knock-on type to drift-diffusion type. This highlights the potential of model pores to illustrate, probe, and quantify ion transport mechanisms. Strain, as well, may help determine the source of $1/f$ noise by modifying mechanical fluctuations and, for large enough pores, only slightly changing other properties (pore size, conductance, etc.). Thus, strain potentially has a variety of uses besides elucidating biological processes and optimizing synthetic pores.

There are other many-body effects primarily relevant at large voltages, such as the polarization concentration we discussed in Sec.~\ref{sec:AccessRes}. Also at larger voltages, water molecules around the pore orient along the applied electric field. This is especially relevant in small (near the dehydration threshold), atomically thin pores due to a substantial drop in voltage across its short channel. In this situation,  ions fluctuating near the pore find ``reoriented'' water molecules that help it enter the pore, increasing the chance of crossing. Such ``polarization induced chaperoning of ions'' contributes to a nonlinear increase in current with voltage~\cite{sahu2017dehydration}. In some 2D membranes, this voltage may come above the degradation threshold for the membrane, but in others, such as MoS$_2$, it may be observable. A similar physical effect can also influence the capture rate of molecules such as DNA~\cite{wilson2018water}. These are just a few examples of the many cases where selectivity, dehydration, and electrostatic interactions all conspire to give a host of complex phenomena.


\section{Key applications}\label{sec:Apps}

Throughout this Colloquium, we have mentioned several applications of ion transport through 2D membranes. Here, we will give additional details of these applications, mainly focusing on the areas that have been extensively studied. The first is filtration, including desalination, molecular sieving, and so on. The second is nanoscale sensing, such as nucleic acid and amino acid sequencing, detection of biochemical specimens, and probing molecular-scale interactions, such as 
dehydration. These applications intend to take advantage of the single atom thickness of 2D material along with other features, such as impermeability, mechanical strength, subnanoscale control in pore formation, and tailorability generally (functionalization, layering, etc.). 

\subsection{Filtration, desalination, and power generation}
Filtration using porous membranes is a mature technology, but 2D membranes may push it to the atomic level, potentially giving new routes to desalination and gas separation. The ultimate goal of filtration is to remove undesirable constituents while maintaining maximum permeation of the filtrates (as with all technologies, the real optimization is to get the final product with minimal resources, e.g., energy, cost, etc.). Achieving these aspects simultaneously is difficult because higher rejection requires stronger interactions with the membrane, which results in the reduction of overall permeation; there is always a trade-off between selectivity and permeation.

Biological pores, such as KcsA, are remarkable since they have large selectivity but still maintain permeation near the diffusion limit. Mimicking this extraordinary design (at scale) with solid-state pores will be a breakthrough in membrane separation technology. Unfortunately, solid-state pores lack the atomic precision of biological pores -- they have surface roughness exceeding the size of hydrated ions~\cite{storm2003fabrication}. A possible alternative is carbon nanotubes whose smooth hydrophobic walls allow fast water flow~\cite{hummer2001water, majumder2005nanoscale}. However, implementation requires large-scale fabrication and uniform directional alignment, which remains challenging~\cite{das2014carbon}.

On the other hand, 2D membranes may allow thousands of pores to be formed in a single sheet, where removal of just a single atom is sufficient to nucleate a pore, which can be enlarged with atomic precision. In fact, these nanopores can be made to selectively filter gases whose size differ only by few tenths of a nanometer~\cite{koenig2012selective, celebi2014ultimate} while, without defects, remain impermeable to even helium~\cite{bunch2008impermeable}. Thus, graphene potentially gives an ideal separation membrane. 

\begin{figure}
\includegraphics{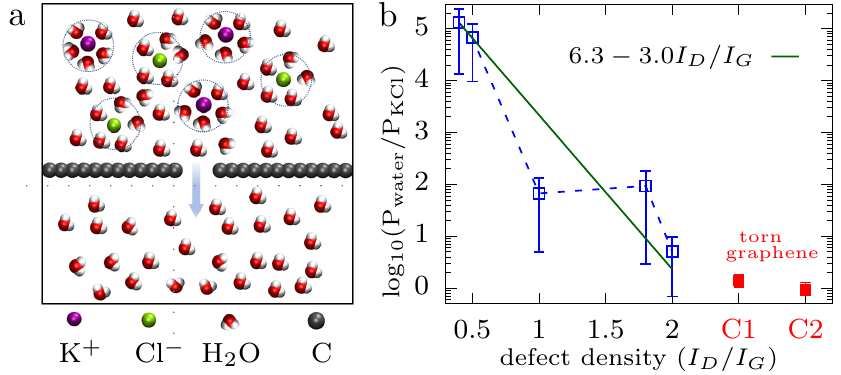}
\caption{Water desalination using a graphene membrane. (a) Schematic of the setup and (b) selectivity of water over salt versus defect density $I_D/I_G$~\cite{Surwade2015}. Due to extended exposure, defect sizes increase with their density. The dashed line is a guide to the eye only.\label{fig:filtration}}
\end{figure}

2D membranes are also attractive for reverse osmosis (RO), which is the industry standard to separate  ions and other solutes from seawater via pressure applied across semipermeable  membranes~\cite{fritzmann2007state}. In this context, 2D membranes can be formed with pores just large enough for water to pass but too small for hydrated ions (and organic molecules) to go through, see Fig.~\ref{fig:filtration}. These pores allow water to flow at an order of magnitude higher rate than commercial RO membranes without compromising the salt rejection~\cite{tanugi2012, Joshi2014, Surwade2015, tanugi2016, ohern2015nanofiltration}. While not yet a complete technology, the high permeation rate may help offset the energy consumption of current RO membranes. Additionally, \citet{rollings2016} suggest that the ability of graphene membranes fabricated under certain conditions to selectivity transport cations over anions may make them useful ion-exchange membrane for electrodialysis. Since the surface charge results in membrane selectivity, the pore size can be much larger than the hydrated ion -- but comparable to the Debye length -- giving a high overall exchange rate.

Other 2D-based membranes, such as GO, are also extensively studied for desalination and filtration. The interest in GO membranes for filtration grew after the observation of fast water vapor transport simultaneous with the blockage of other species, including helium~\cite{nair2012}. Such membranes were later demonstrated to block ions~\cite{Joshi2014}, including controllable channel height~\cite{abraham2017}, see Figs.~\ref{fig:selectivity}b-d. Other approaches control the GO interlayer spacing with cations~\cite{chen2017ion} or use alternate structures with 2D materials as spacers~\cite{esfandiar2017size} to create a nanoscale slit for water passage. GO membranes have likewise been used for hydrogen separation~\cite{li2013} and water removal from organic solvents~\cite{yang2017ultrathin}. Furthermore, \citet{ji2017osmotic} recently demonstrated that negatively charged GO membranes (n-GO) in  pristine form can be functionalized to make them positively charged (p-GO), thus providing both cation-selective and anion-selective membranes.

From the energy perspective, reverse osmosis and electrodialysis processes convert pressure and electrical energy to electrochemical gradient energy. Conversely, pressure retarded osmosis (PRO) and reverse electrodialysis (RED) generate power from a salinity gradient across the membrane~\cite{logan2012membrane}.  In one form or another, most potential membrane applications make use of reducing the resistance to flow~\cite{geim2011nobel}, while maintaining selectivity. To this end, 2D materials give promising semipermeable or ion-selective membranes for osmotic power generation~\cite{siria2017new}. An ideal membrane for PRO should allow the fast flow of water and be able to withstand a large pressure difference~\cite{siria2017new}. 2D membranes demonstratively excel in the first requirement; and, although they will require an additional support layer to withstand the high-pressure difference, their exceptional strength will allow overall thickness to be small and pore sizes to be larger than traditional membranes~\cite{gai2014ultrafast}.  On the other hand, RED does not require the membrane to withstand very high pressure; additionally, electric current is directly generated from this process without the requirement of turbines. Experiments have demonstrated the potential of single-layer MoS$_2$, hBN, and graphene membranes as ion-exchange membrane for power generation using a concentration gradient across a selective membrane to drive a current~\cite{feng2016single, walker2017}. Since the ion selectivity is due to membrane charge, it should persist even for large pores. This result in very high power density of 1 MW/m$^2$ in MoS$_2$ ~\cite{feng2016single} assuming uniform pores of 10 nm diameter with membrane porosity 30\%. \citet{walker2017} estimated power density of  0.7 kW/m$^2$ in graphene and hBN with multiple pores of diameter around 1 nm. These estimates -- even when lowering the porosity or other factors to more realistic values -- are impressive compare to the power density of 0.5 W/m$^2$  in commercial ion exchange membrane and 0.77 W/m$^2$ in thicker GO membranes~\cite{ji2017osmotic}. Boron nitride nanotubes have also shown to have very high power density, 4 kW/m$^2$ ~\cite{siria2013giant}, on the open surface of the tube. The macroscopic power density will, however, depend on the packing density. Also, the large-scale production and alignment of nanotubes is a challenge~\cite{siria2017new}.  

As with nanotubes (whether carbon or boron-nitride), it remains a challenge to develop techniques amenable to industrial scale fabrication. Any method needs to be both inexpensive and scalable. Drilling pores one by one using an ion or electron beam would be too costly. Methods such as dielectric breakdown~\cite{kwok2014nanopore,kuan2015electrical} are less expensive, as they do not require expensive instruments, and may provide a solution. Other approaches are being examined or suggested, such as broad ion-beam exposure of a particular intensity and voltage~\cite{russo2012atom}, ion bombardment followed by chemical etching~\cite{OHern2014, ohern2015nanofiltration}, or ozone treatment~\cite{walker2017}.  To our knowledge, none of these approaches have yet been demonstrated to have commercial implications.

In addition to the fabrication challenges, there are other roadblocks, such as performance deterioration due to fouling, i.e., blockage of the pores by deposition of biological and chemical impurities. The membrane themselves can also break down over time. These problems are inevitable, but some materials are less susceptible. Encouraging studies have shown that graphene-based materials have antibacterial~\cite{hu2010graphene, liu2011antibacterial} and antifouling~\cite{lee2013graphene, seo2018anti} properties, as well as improved durability~\cite{choi2013layer}. If these transfer to a final technology, this could reduce costs by increasing the lifetime of RO membranes and reducing regular maintenance requirements. 

\subsection{Molecular sensing and sequencing}
The idea of using ion transport through pores to detect specimens dates back half a century. The Coulter counter~\citep{coulter1953means} is a well-known approach that counts and sizes a particle by observing a pulse in ionic current caused by its translocation. The nanopore sequencers from Sec.~\ref{sec:Classes} are molecular-scale Coulter counters that detect via the ionic blockade of a specimen translocating through the pore. In some sense, molecular detection can trace its genesis back to this development. 

Early research in next-generation sequencing  aimed at reducing the cost and time of sequencing. When the first entire human genome was completed in 2003, it involved 13 years of collaboration between hundreds of scientists and a \$3 billion cost~\cite{collins2003human}. The price and time required have decreased rapidly since then; sequencing can now be done within a few days for about \$1000 per genome~\cite{hayden2014technology}. This is a remarkable feat, but it comes with the drawbacks of higher error rates and shorter read length~\cite{quail2012tale, goodwin2016coming}. The equipment necessary for sequencing is still expensive, costing several million dollars. Nanopore sequencing, in contrast, may further reduce the cost and time, make the process portable for on-site sequencing (e.g., point-of-care diagnostics), and give sensors that may be embedded in other devices. These will be enabled by minimal sample preparation, being label-free, and ability to be integrated into electronic circuits for signal processing and communication~\cite{branton2008potential}. As we discussed in Sec.~\ref{Sec:Pores}, sensing using nanopores has come a long way from mere detecting DNA translocation. In addition to sequencing, other sensing applications using nanopores are emerging, such as detection of protein folding~\cite{si2017nanopore}, protein analytes~\cite{movileanu2000detecting}, peptides~\cite{chavis2017single}, virus particles~\cite{yang2016quantification}, and cancer-markers~\cite{duan2018label}. Protein sequencing is another promising application of nanopores as shown by several studies~\cite{wilson2016graphene, kolmogorov2017single, farimaniidentification2018}. 

The success of nanopore sequencing has primarily been associated with the use of biological pores. Even so, solid-state technology could provide considerable benefits for molecular detection, such as the ability to operate at extreme temperature, voltage, and \pH. The primary reason for the success of biological pores is the atomically precise structure, modularity, and size of their sensing region. In order for artificial pores to sequence competitively, they need to have the same level of atomic control and precision. 2D materials provide such an opportunity since the sensing regime (whether for blockade or transverse transport) is the size of a single nucleotide. 

While several groups measured ionic blockade from DNA translocation through graphene, MoS$_2$, and hBN pores, several challenges -- such as reducing translocation speed, preventing DNA from sticking to the surface, controlled feeding of DNA into the pore, and reducing $1/f$ noise -- need to be overcome before individual base discrimination in a DNA will be possible. Progress is being made to address some of these challenges. The longer residence time in the sensing region, for instance, is particularly important to give a better SNR. \citet{wells2012}, for instance, suggested the possibility of using the binding of DNA to graphene to slowly feed it through the pore. This is yet to be realized in experiments.  \citet{feng2015identification} were able to reduce the translocation speed through MoS$_2$ nanopores using a viscous ionic liquid on one side of the pore. This enabled them to distinguish between homopolymers of different DNA bases and even individual isolated nucleotides. Functionalization of the pore with groups that can hydrogen bond with DNA is also a possibility (or, in general, binding to increase dwell time and distinguishability of analytes). In the case of graphene, though, we agree with the assessment of \citet{heerema2016} that ion transport blockades alone likely will not be a successful approach due to the many issues discussed above.

\begin{figure}
\includegraphics{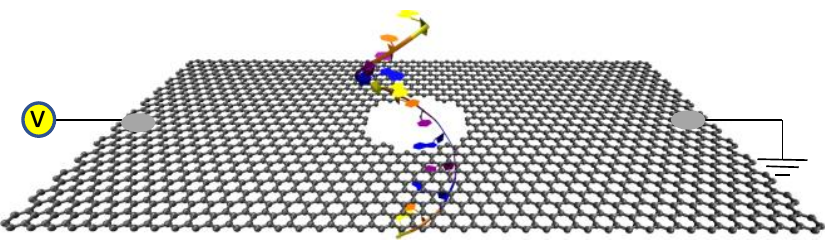}
\caption{Schematic of DNA sequencing via the transverse current. As a DNA  translocates electrophoretically (or by other means), the nucleotide in the pore modulates the in-plane current through the graphene, identifying the base present.\label{fig:transverse}}
\end{figure}

Sensing with nanopores can also potentially be done via other modalities, such as the transverse tunneling or in-plane current, Fig.~\ref{fig:transverse}. Proof-of-principle of sequencing via transverse transport~\cite{Zwolak05,Zwolak08, lagerqvist2006fast,lagerqvist2007comment,krems2009effect, lagerqvist2007influence} has already been demonstrated~\cite{ tsutsui2010identifying, chang2010electronic, tsutsui2011electrical, huang2010identifying, tsutsui2012transverse}.  Graphene, in particular though, can act as both membrane and electrode~\cite{heerema2016}, see Fig.~\ref{fig:transverse}, thus reducing the complexity of the device. Such sensing or sequencing may be realized by using tunneling through a graphene nanogap~\cite{postma2010rapid} or modulation of the in-plane current due to the presence of DNA base in the nanopore~\cite{nelson2010detection} or due to adsorption of DNA on a graphene ribbon~\cite{heerema2016, song2012nucleobase, min2011fast}. Additionally, both the ionic and in-plane currents can be measured simultaneously~\cite{traversi2013detecting,heerema2018probing}. Comparing the two signals can filter out noise and identify correlated events, yielding more information on DNA translocation. Also, since there can be strong capacitive effects~\cite{balijepalli2014} in a typical nanopore setup, others~\cite{lathrop2010monitoring, sigalov2008detection} have shown that the alternating current, in addition to the direct current, can be used to control and analyze DNA translocation through a pore. Regardless of modality (ion transport, in-plane or tunneling electronic transport, etc.), the development of a  successful device is likely to be a concerted effort by many groups and fields and will require a highly integrated device, with ``on-chip'' control and amplification. 

There are also alternative applications of transverse currents. One can detect electrostatic fluctuations and sense changes in protein structure and other biochemical phenomena. This is well covered in other reviews~\cite{allen2007carbon,stine2012fabrication}. Deflection of graphene due to (bio)molecular binding to functional groups, protein unfolding through a pore, and other processes may also give a method of detection~\cite{gruss2017communication,gruss2018graphene}. Deflection stretches covalent bonds, weakening them, and thereby reducing the electronic current in the graphene sheet. This was originally suggested as a route to sequencing~\cite{paulechka2016nucleobase,smolyanitsky2016mos2}. However, various sources of noise and errors, such as false positives (adsorption of DNA or just steric interactions) and electrostatic interactions (local gating also changes currents), as well as the factors that hinder nanopore sequencing in general (lack of control of DNA motion, etc.),\ give substantial obstacles. Graphene deflectometry, though, may be useful in detecting weak and fast molecular-scale forces, in some cases requiring an appropriate, specialized assay -- as in atomic force microscopy studies of biomolecules -- that functionalization of graphene would enable~\cite{gruss2017communication,gruss2018graphene}.


\section{Conclusions and future directions}\label{sec:conc}

There are two sides of the 2D membrane: fundamentals and applications. We have seen both, as well as their synergy, in this Colloquium. For instance, porous membranes and channels in heterostructures and GO laminates give a platform to study the competition between dehydration and interaction with functional groups, as well as other characteristics such as reduced mobility. These fundamental aspects of ion transport determine selectivity and influence molecular and water transport. Moreover, as \citet{abraham2017} argue, GO membranes may be manufacturable at the industrial scale, giving a potential route to commercial membranes with high water flow rates and selectivity. Scaling up is still challenging~\cite{werber2016materials}. It is clear, then, that a pressing direction is to develop techniques that are amenable to the bulk manufacturing of membranes with desirable transport characteristics. A related issue is to characterize the resulting membranes, the functional groups present, the role of adsorbed species (e.g., in creating membrane charge), edge composition, etc., and how these factors change with the fabrication process and conditions.

On the fundamental side, porous membranes, especially single well-controlled pores, potentially give ideal models for understanding aspects of biological ion channels. Heterostructures and GO laminates also provide such opportunities. Single pores, in particular, have a structural similarity to the selectivity filter and lack ensemble averaging (i.e., translocating ions do not pass by many functional groups/adsorbed species). Just as above, one of the significant issues is characterization -- to determine the atomic structure of the pore edge and membrane composition. Another problem is control -- one needs to know precisely what is there and be able to change it. Selective functionalization is challenging and a nascent area with a vast potential for impact. For individual pores the radius and length (via layering) is reasonably well controllable but still with some limitation. Putting all these components together -- control of diameter, thickness, inhomogeneous layering, functional composition -- will enable a broad and systematic study of ion transport at the confluence of different energy and length scales.

These considerations also make simulation and interpreting experiments challenging: Transport through nanoscale pores in 2D membranes has contributions from both pore and access resistance; is typically in a regime where the Debye length is comparable to the membrane thickness and pore diameter; is influenced by dehydration and interactions; and has ill-defined basic parameters such as radius and thickness. Moreover, the lack of ensemble averaging means that each pore may be different --  unknown functional groups/charges can hinder comparisons of theory and experiment. While the latter is an issue across many fields in nanoscale science, there is now a route to tackle the others. \citet{sahu2018golden,sahu2018maxwell} showed, for instance, that a scaling ansatz and the “golden aspect ratio” captures both the pore and the access resistance, yielding a simple method with low computational cost. This is in addition to the routine care required for accurate simulation (long simulations that reduce statistical errors, simulation cells that do not give cross-talk with periodic images, and proper quantification of the potential drop). The development of accurate, polarizable force fields (especially for graphene) and ab initio MD simulations will refine our estimates for the energetics of ion transport through subnanoscale pores, where dehydration and ion--membrane interactions (including with charges, but also the electrons in the membrane), are essential. These will help to quantitatively assess mechanisms in biomimetic pores, as well as in sensors that rely on electrostatic gating. An orthogonal question regards the behavior of flow fields when standard approximations (extended channels and equilibrium distributions transverse to the direction of motion~\cite{schoch2008transport}), cannot be made, and when there is a back action of the fluctuating membrane on water/ion motion. 

We started this Colloquium with an overview of biological channels and the development of nanopore-based sequencing. Will graphene or other 2D membranes overtake their biological counterparts in sequencing and sensing technologies? Only time will tell. While their atomic thickness (e.g., a naturally high spatial resolution), and stability confer significant advantages, ion and molecular transport still suffer from drawbacks due to rapid translocation and interactions. These materials offer the opportunity to create integrated electronic sensing that may make headway into sequencing and other sensing technologies. However, pores and channels made with 2D membranes also provide something else entirely: a chance to create simplified versions of biological channels -- a kind of ``bio-lite'' ion channel. These will enable the delineation of dehydration, interactions, static structure, and fluctuations. Overall, this will push our knowledge and understanding of nanoscale ion transport to new heights, allowing for discoveries in such diverse fields as drug design, simulation, and filtration, among many others.  

\acknowledgments
We are grateful to J. Elenewski, M. Ochoa, C. Rohmann, J. P. Zwolak, J. A. Liddle, S. Stavis, K. Briggs, V. Tabard-Cossa, and J. Kasianowicz for providing helpful comments. S. S. acknowledges support under the Cooperative Research Agreement between the University of Maryland and the National Institute of Standards and Technology Center for Nanoscale Science and Technology, Award 70NANB14H209, through the University of Maryland.


%

\end{document}


\title{Supplementary Material: Ionic phenomena in nanoscale pores through 2D materials}
\author{Subin Sahu}
\affiliation{\mbox{Biophysics Group, Microsystems and Nanotechnology Division, Physical Measurement Laboratory}, \mbox{National Institute of Standards and Technology, Gaithersburg, Maryland 20899, USA}}
\affiliation{\mbox{Maryland NanoCenter, University of Maryland, College Park, Maryland 20742, USA}}
\author{Michael Zwolak}
\email[\textbf{Corresponding author:} ]{mpz@nist.gov}
\affiliation{\mbox{Biophysics Group, Microsystems and Nanotechnology Division, Physical Measurement Laboratory}, \mbox{National Institute of Standards and Technology, Gaithersburg, Maryland 20899, USA}}

\maketitle
\tableofcontents

\clearpage

\section{Conductance versus pore radius from experiment}

\begin{figure}[h]
\includegraphics{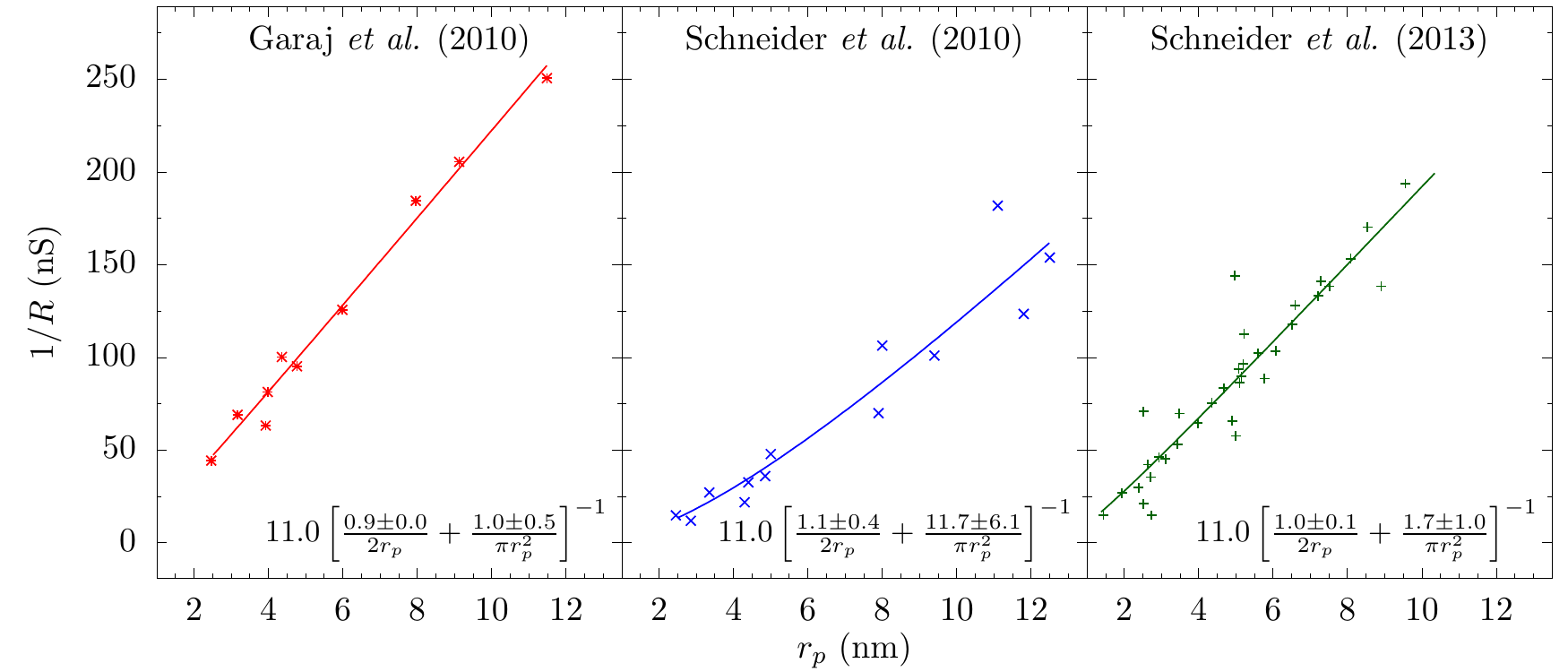}
\caption{Conductance fits for experimental data from \citet{garaj2010}, \citet{Schneider2010}, and \citet{schneider2013}. Note that the fits here are slightly different than ones given in the respective papers, as indicated in the main text. However, the basic conclusion is the same: \citet{garaj2010} and \citet{schneider2013} obtained a dominant access contribution, whereas \citet{Schneider2010} obtained a dominant pore contribution. The pore radii in these experiments are measured using the TEM image of the pores which roughly corresponds to $r_p$ in Fig.~7 of the main text -- i.e., the average distance from the pore center to the pore edge excluding the carbon electron cloud (approximately its vdW radius). However, for the size of the pores presented here, the difference between the various definitions of pore radius is not significant (similarly for the exact pore geometry).}
\end{figure}

\clearpage

\section{Conductance versus pore radius from all-atom molecular dynamics}

\begin{figure}[h]
\includegraphics{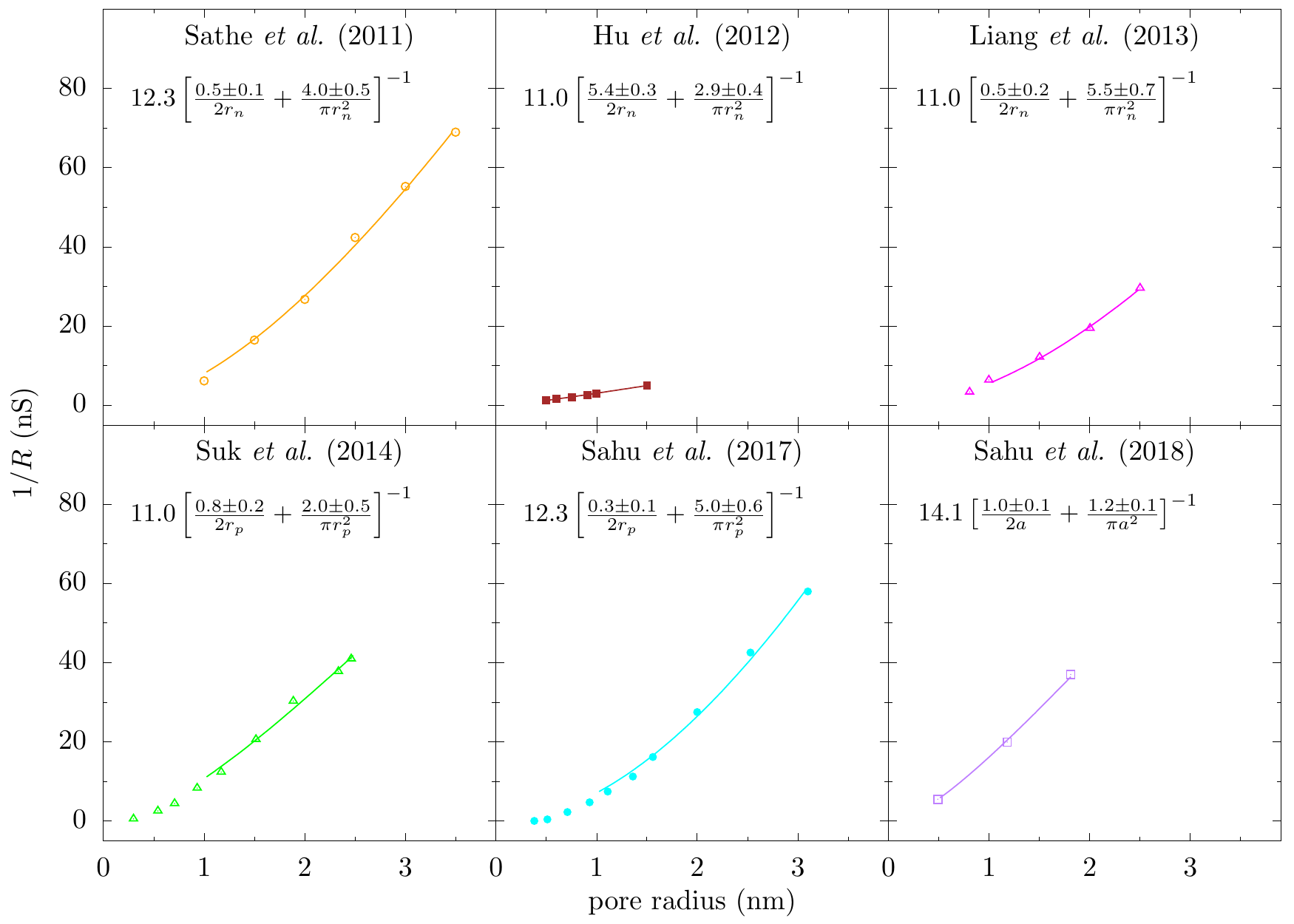}
\caption{Conductance fits for all-atom MD data from \citet{sathe2011}, \citet{Hu2012}, \citet{liang2013theoretical}, \citet{suk2014ion}, \citet{Sahu2017dehydration}, and \citet{sahu2018maxwell}. The definition of pore radius varies in these studies. \citet{sathe2011}, \citet{Hu2012}, and \citet{liang2013theoretical} take the pore radius as the nominal radius employed in cutting the pore -- i.e., all carbon atoms with coordinates that satisfied $r_n^2 \le x^2+y^2$ were removed to create the pore and $r_n$ was taken as pore radius (note that $r_n$ is not unique with this definition, a range of $r_n$ give the same physical pore. We use the values as given in th respective papers). \citet{suk2014ion} take the pore radius as the distance from the pore center to where the water density dropped below 2 \% of the bulk water density. \citet{Sahu2017dehydration} define the pore radius as the average distance from the center of the pore to the inner edge of the pore atoms, i.e., $r_p$ as shown with red dashed line in Fig.~7 of the main text. Both the \citet{suk2014ion} and \citet{Sahu2017dehydration} definitions give similar values for the pore radius. However, \citet{sahu2018maxwell} define the pore radius from the current density profile in the pore, as shown by $a$ in Fig.~7 of the main text. Solid lines are shown where data was fit. We only fit the data in which the pore is large enough to ignore the effects of hydration, except in the case of \citet{Hu2012}, which only has small pores. We note that dehydration increases the pore resistance drastically and thus overshadows the access resistance. For instance, in the case of \citet{Sahu2017dehydration}, only the dehydration regime was of interest. Error bars were not considered in the fits since not all the data were reported with error bars (in particular, the ones in the top panel). }
\end{figure}

\newpage
\section{Physical/chemical properties and applications}

\begin{table}[h]
    \centering
    \def\arraystretch{1.2}
\setlength\tabcolsep{8 pt}
    \begin{tabular}{c| c c c}

           & Graphene & hBN  & MoS$_2$ \\ \hline  \hline
    Thickness (vdW) &   0.335 nm  & 0.334 nm & 0.65 nm  \\ \hline
    Young's modulus &  1 TPa (130 
    GPa) &   0.27 TPa (23 GPa)  & 0.865 TPa (70.5 GPa)  \\
        (Breaking strength)           & \cite{lee2008measurement}       &  \cite{bertolazzi2011stretching}  & \citet{falin2017mechanical}\\ \hline 
    
    Wettability & Hydrophobic  & Mild hydrophilic & Hydrophilic \\
      & \cite{schneider2013}  &  \cite{zhou2013dna}  & \cite{liu2013top} \\ \hline
    Surface charge & $-0.6$ C/m$^2$(\pH 8)    & $-0.07$C/m$^2$ to $-0.16$ C/m$^2$(\pH 7)   & $-0.024$ C/m$^2$ to $-0.088$ C/m$^2$        \\ 
    density & \cite{rollings2016} &  \cite{weber2017boron} &  (\pH 5)  \\
    &  -0.039 C/m$^2$(\pH 5)   & &\cite{feng2016single}        \\ 
     & \cite{ shan2013surface} &  &  \\\hline
    \end{tabular}
    \caption{Comparison of the physical/chemical properties 2D materials relevant to topics covered in the main text. We list here the atomic thickness -- defined including the van der Waals radii -- for a simple comparison. The effective thickness for ion transport can be significantly larger than the thicknesses in this table and it depends on, for example, voltage (see the main text). The large value of Young's modulus and breaking strength demonstrates the extraordinary strength of 2D materials. For comparison, the Young's modulus and breaking strength of steel are 200 GPa and 500 MPa, respectively~\cite{ledbetter1980elastic}. The hydrophobicity/hydrophilicity listed in the table are for the nanopore in the membrane rather than the flat surface of the membrane; thus, although MoS$_2$ membrane is hydrophobic, the pore is hydrophilic due to Mo-rich edges. The surface charge densities listed here are just a few representative examples as it can vary widely based on the fabrication technique (e.g., exfoliated versus chemical vapor deposition growth of the material), defect density, and other environmental factors (\pH, for instance). Additionally, the charge density for porous membrane can be different than that for a pristine membrane. As a point of comparison, the surface charge for silicon nitride is typically reported to be around 20 mC/m$^2$ to 60 mC/m$^2$ at \pH 7 to \pH 8~\cite{ho2005electrolytic,smeets2006salt}.}
    \label{tab:my_label}
\end{table}

\begin{table}[h]
    \centering
    \def\arraystretch{1.2}
\setlength\tabcolsep{8 pt}
    \begin{tabular}{|c| c|}
    \hline
    Graphene &  Gas separation~\cite{celebi2014ultimate, koenig2012selective},\\   & Osmotic power \cite{walker2017, gai2014ultrafast}, \\
    & Desalination\cite{tanugi2012,  Surwade2015, rollings2016}, \\
    & Sensing and sequencing \cite{garaj2010, merchant2010, Schneider2010}  \\
    & \cite{sathe2011, wells2012, heerema2016} \\ \hline
    GO & Desalinaiton \cite{chen2017ion, nair2012, Joshi2014, abraham2017},\\
    & Gas separation~\cite{li2013}, Solvent recovery \cite{yang2017ultrathin},\\
    &  Osmotic power \cite{ji2017osmotic}    \\ \hline
    MoS$_2$ & Desalination \cite{hirunpinyopas2017desalination}, Osmotic power \cite{feng2016single} \\
    & Sensing and sequencing \cite{liu2014atomically, feng2015identification}\\
     \hline
    hBN  & Sensing and sequencing \cite{liu2013boron, zhou2013dna}    \\ 
    & Osmotic power \cite{walker2017}    \\ \hline
    2D-heterostructures  & Desalination \cite{esfandiar2017size, gopinadhan2019complete}    \\ \hline
   Biomimetic \&  & Ion channel mechanisms~\cite{sahuOptimal2019} and biomimetics~\cite{sint2008, he2013bioinspired}, \\
   fundamental  &  Ionic coulomb blockade~\cite{feng2016observation}  \\ 
   studies & Dehydration (exp)~\cite{Joshi2014, abraham2017, esfandiar2017size} \\ 
    & Dehydration (th)~\cite{ Sahu2017dehydration, sahu2017ionic} \\ \hline
    \end{tabular}
    \caption{Various applications of 2D material and their derivatives. This list is not exhaustive. It includes the papers we discussed in the main text but these applications are often touched on in many studies and are part of larger and very active research fields. Many fundamental studies have their origin in related work with traditional solid-state pores, see the main text and the references in the citations above.}
    \label{tab:my_label}
\end{table}
\newpage


%